\documentclass[prd,preprintnumbers,floatfix,
nofootinbib,superscriptaddress]{revtex4}

\usepackage{float}
\usepackage{nicefrac}
\usepackage{mathtools}
\usepackage{amsfonts} % AMS
\usepackage{amssymb} % AMS
\usepackage{amsmath} % AMS
\usepackage{graphicx} % Include figure files
\usepackage{subfigure} % Include figure files
\usepackage{array} % array
\usepackage{dcolumn} % Align table columns on decimal point
\usepackage{bm} % bold math
\usepackage{latexsym} % latex symbols
\usepackage{longtable} % long tables
\usepackage{hyperref} % hypertext links 
\usepackage{verbatim}
\usepackage{epsfig}
\usepackage{slashed}
\usepackage{color}
\renewcommand{\vec}[1]{\mathbf{#1}}
\usepackage{tikz}
\DeclareRobustCommand\sampleline[1]{%
  \tikz\draw[#1] (0,0) (0,\the\dimexpr\fontdimen22\textfont2\relax)
  -- (2em,\the\dimexpr\fontdimen22\textfont2\relax);%
}

\newcommand{\Kdf}[0]{{\cK_{\df,3}}}
\newcommand{\bp}[0]{\mathbf{p}}

\newcommand{\cD}[0]{\mathcal D}
\newcommand{\cK}[0]{\mathcal K}
\newcommand{\cL}[0]{\mathcal L}
\newcommand{\cM}[0]{\mathcal M}
\newcommand{\cO}[0]{\mathcal O}
\newcommand{\cR}[0]{\mathcal R}
\newcommand{\cS}[0]{\mathcal S}
\newcommand{\cT}[0]{\mathcal T}

\newcommand{\one}[0]{{(1)}}

\newcommand{\df}[0]{\mathrm{df}}
\newcommand{\K}[0]{\mathcal K}

\newcommand{\uu}[0]{{{(u,u)}}}

\newcommand{\bhat}[1]{\widehat{\mathbf{#1}}}
\newcommand{\THrho}[0]{[\Theta \bar \rho]}

\newcommand{\BHSQC}[0]{Briceno:2017tce}
\newcommand{\BHSnum}[0]{Briceno:2018mlh}
\newcommand{\HSQCa}[0]{Hansen:2014eka}
\newcommand{\HSQCb}[0]{Hansen:2015zga}

\newcommand{\nn}[0]{\nonumber}

\begin{document}

\preprint{\vbox{\hbox{JLAB-THY-19-2945} }}
\preprint{\vbox{\hbox{CERN-TH-2019-078} }}

\title{Unitarity of the infinite-volume three-particle scattering amplitude\\ arising from a finite-volume formalism}

%%%%%%%%%%
\author{Ra\'ul A. Brice\~no}
\email[e-mail: ]{rbriceno@jlab.org}
\affiliation{Thomas Jefferson National Accelerator Facility, 12000 Jefferson Avenue, Newport News, VA 23606, USA}
\affiliation{ Department of Physics, Old Dominion University, Norfolk, Virginia 23529, USA}
%%%%%%%%%%

%%%%%%%%%%
\author{Maxwell T. Hansen}
\email[e-mail: ]{maxwell.hansen@cern.ch}
\affiliation{Theoretical Physics Department, CERN, 1211 Geneva 23, Switzerland}
%%%%%%%%%%

%%%%%%%%%%
\author{Stephen R. Sharpe}
\email[e-mail: ]{srsharpe@uw.edu}
\affiliation{Physics Department, University of Washington, Seattle, WA 98195-1560, USA}
%%%%%%%%%%

%%%%%%%%%%
\author{Adam P. Szczepaniak}
\email[e-mail: ]{aszczepa@indiana.edu}
\affiliation{Physics Department, Indiana University, Bloomington, IN 47405, USA}
\affiliation{Center for Exploration of Energy and Matter, Indiana University, Bloomington, IN 47403, USA}
\affiliation{Thomas Jefferson National Accelerator Facility, 12000 Jefferson Avenue, Newport News, VA 23606, USA}

%%%%%%%%%%

%%%%%%%%%%
\date{\today}
%%%%%%%%%%

%%%%%%%%%%
\begin{abstract}
In Ref.~\cite{\HSQCb}, two of us derived a relation between the scattering amplitude of three identical bosons, $\mathcal M_3$, and a real function referred to as the {divergence-free} K matrix and denoted $\Kdf$. The result arose in the context of a relation between finite-volume energies and $\Kdf$, derived to all orders in the perturbative expansion of a generic low-energy effective field theory. 
In this work we set aside the role of the finite volume and focus on the infinite-volume relation between $\Kdf$ and $\mathcal M_3$. We show that, for any real choice of $\Kdf$, $\mathcal M_3$ satisfies the three-particle unitarity constraint to all orders. Given that $\Kdf$ is also free of a class of kinematic divergences, the function may provide a useful tool for parametrizing three-body scattering data. 
Applications include the phenomenological analysis of experimental data (where the connection to the finite volume is irrelevant) as well as calculations in lattice quantum chromodynamics (where the volume plays a key role).
 \end{abstract}
%%%%%%%%%%

%\keywords{weak decays, lattice QCD}

\nopagebreak

\maketitle

%%%%%%%%%%%%%%%%%%%%%%%%%%%%%%%%%%%%%%%
%%%%%%%%%                          PAPER                             %%%%%%%%%
%%%%%%%%%%%%%%%%%%%%%%%%%%%%%%%%%%%%%%%

\section{Introduction\label{sec:intro}}

Three-body systems lie at the forefront of modern-day theoretical hadronic physics. Whether in the context of understanding the resonance spectrum of quantum chromodynamics (QCD) or the binding of nucleons in nuclei, three-body dynamics play a crucial role. In recent years there has been significant progress  
in developing rigorous theoretical frameworks for studying such systems. 

The majority of QCD states are unstable resonances that decay via the strong force into multihadron configurations. A quantitative description of these is given by identifying complex-valued energy poles in the scattering amplitudes of the resonance decay products. Given that one can only access real-valued energies experimentally, it is necessary to construct amplitude parametrizations that can be analytically continued into the complex energy plane, in order to determine the pole positions. Since resonance widths originate from the presence of open decay channels, 
unitarity plays a key role in the analytic continuation. 
It is straightforward to impose 
 unitarity  on  two-body amplitudes, but it far more challenging in the 
  three-body case, with efforts dating back to the 1960s~\cite{Bjorken:1960zz, Aitchison:1966lpz, Ascoli:1975mn}. 
  
The availability of high-precision data on various three-body production and resonance decay channels, together with the emergence of lattice QCD (LQCD) calculations of hadron scattering, has reignited interest in the three-body problem~\cite{Magalhaes:2011sh, Mai:2017vot, Mikhasenko:2019vhk, Jackura:2018xnx}. 
Although unitarity  gives a powerful restriction on the structure of scattering amplitudes, it does not 
fully determine them. 
 The unconstrained real part, 
 often referred to as the K matrix,  
 is determined by the underlying microscopic theory,
   and in practice is obtained by fitting to experimental data or LQCD finite-volume spectra. 
   By comparing results obtained with different K-matrix  parametrizations it is possible to determine the existence of amplitude singularities and learn about their microscopic origin. 
  This approach has proven remarkably powerful, not only for the determination of simple QCD observables, but also in multiparticle quantities including scattering and transition amplitudes. 
 
 In LQCD, using the standard approach, one can directly access only
 the eigenstates and energies of the finite-volume Hamiltonian, 
 which are not in direct correspondence to multiparticle asymptotic states. 
  This prevents a direct determination of S-matrix elements. 
  Nevertheless, it turns out that one can extract scattering information via model-independent relations between finite- and infinite-volume quantities.
For two-particle systems there has been a great deal of progress in developing such formalism, culminating in a general relation between the finite-volume spectrum of any coupled two-particle system and its corresponding scattering matrix~\cite{Luscher:1986n2,Luscher:1991n1,Rummukainen:1995vs,Kim:2005gf, He:2005ey, Lage:2009,Fu:2011,Hansen:2012tf, Briceno:2012yi,Gockeler2012,Briceno:2014oea}. In addition, relations have been derived between finite-volume matrix elements and the corresponding transition amplitudes mediated by an external current~\cite{Lellouch:2000,Christ:2005, Meyer:2011um,Bernard:2012bi, Agadjanov:2014kha,Briceno:2014uqa, Feng:2014gba, Briceno:2015csa, Briceno:2015tza, Baroni:2018iau}. These relations, along with algorithmic advances, have made possible the study of resonant and non-resonant scattering amplitudes of various two-body channels~\cite{Wilson:2015dqa, Briceno:2016mjc, Brett:2018jqw, Guo:2018zss, Andersen:2017una, Andersen:2018mau} including energies where more than one channel is open~\cite{Dudek:2014qha, Dudek:2016cru, Woss:2018irj, Woss:2019hse}. We point the reader to Ref.~\cite{Briceno:2017max} for a recent review of the formalism and its implementation.

Presently, the extension of these studies to energies above 
  three-particle thresholds is limited as the required three-body finite-volume formalism is still under development, although finite-volume energy levels coupling to three-particle states are already 
   being extracted using lattice QCD~\cite{Beane:2012vq, Cheung:2017tnt, Woss:2019hse, Detmold:2013gua, Horz:2019rrn}. The need for this extension has motivated several efforts~\cite{Polejaeva:2012ut,Briceno:2012rv, \HSQCa,\HSQCb,Hammer:2017uqm,Hammer:2017kms,Guo:2017ism,Mai:2017bge, \BHSQC, Doring:2018xxx, \BHSnum, Mai:2018djl, Briceno:2018aml, Blanton:2019igq}, which were recently reviewed in Ref.~\cite{Hansen:2019nir}. At this stage, the formal approach is complete for  systems of three identical scalar particles, including systems with two-to-three transitions as well as those with a resonant two-particle subprocess. 

In this article we restrict attention to the formalism introduced by two of us in Refs.~\cite{\HSQCa,\HSQCb}. This approach, derived via an all-orders perturbative expansion of a generic scalar field theory, relates finite-volume energy levels to an intermediate infinite-volume quantity referred to as the three-body divergence-free K matrix, and denoted $\mathcal K_{\text{df},3}$. In a second step this real-valued intermediate quantity is related, using a set of known integral equations, to the complex valued three-to-three scattering amplitude, $\mathcal M_3$.
Qualitatively, one can understand $\mathcal K_{\text{df},3}$ as the 
part of the scattering amplitude that
 describes all of the microscopic interactions between the three 
 particles that remain 
  after the explicit effects of particle exchanges are subtracted.  
This is somewhat analogous to the relation between the real-valued K matrix and the complex scattering amplitude in the two-particle sector, reviewed in Sec.~\ref{sec:scattering} below. 

In this work we set aside the role of the finite-volume and consider the implications of the relation between $\Kdf$ and $\cM_3$. We demonstrate, to all orders in a $\mathcal K_{\text{df},3}$ expansion, that any scattering amplitude expressed in terms of this real-valued quantity exactly satisfies three body unitarity. 
We stress here that the formulation is fully relativistic and incorporates all partial waves in the three-particle system as well as its two-particle subsystems. We do, however, restrict attention to the relations of Ref.~\cite{\HSQCb}, meaning that the expressions describe a single channel of three identical scalars.

We stress that our result is expected,
since the derivation of the expression for $\cM_3$
in terms of $\Kdf$ is based on an all-orders analysis in quantum field theory.
 Nevertheless, since the derivation is complicated and lengthy, our result provides
an important cross check of the final expression. In addition, we hope that our result stimulates
comparison of the unitary expression for $\cM_3$
in terms of $\Kdf$ with other unitary parametrizations, 
such as that of Refs.~\cite{Mai:2017vot,Jackura:2018xnx}.

The remainder of this work is organized as follows. 
In Sec.~\ref{sec:scattering}, in addition to introducing some basic notation, we review the definition of the scattering amplitude in terms of the K matrix in both the two- and three-particle sectors.
Next, in Sec.~\ref{sec:unitarity}, we review the unitarity relation, with some details relegated to the appendix, and demonstrate that $\mathcal M_3[\mathcal K_{\text{df},3}]$ exactly satisfies the constraining equation. The derivation proceeds in two steps, first showing that the relation holds for $\Kdf=0$ and then incorporating the all-orders effects of the local three-body interaction. 
We conclude briefly in Sec.~\ref{sec:conclusion}.

\section{Two- and three-body scattering\label{sec:scattering}}

In this section we set up some of the notation and key relations used in this work to describe both two- and three-particle scattering. First, in the following subsection, we introduce the two-particle scattering amplitude and recall how its relation to the K matrix automatically satisfies unitarity. Then, in Sec.~\ref{sec:3scatt}, we give the relation between the fully-connected three-particle scattering amplitude, $\mathcal M_3$, and $\Kdf$. In this case both the unitarity constraint and the relation between K matrix and scattering amplitude are more complicated. However, as we show in Sec.~\ref{sec:unitarity}, any form of $\mathcal M_3$ defined in terms of a real-valued $\Kdf$ will satisfy the unitarity constraint.

\begin{figure}
\begin{center}
\includegraphics[width=.7\textwidth]{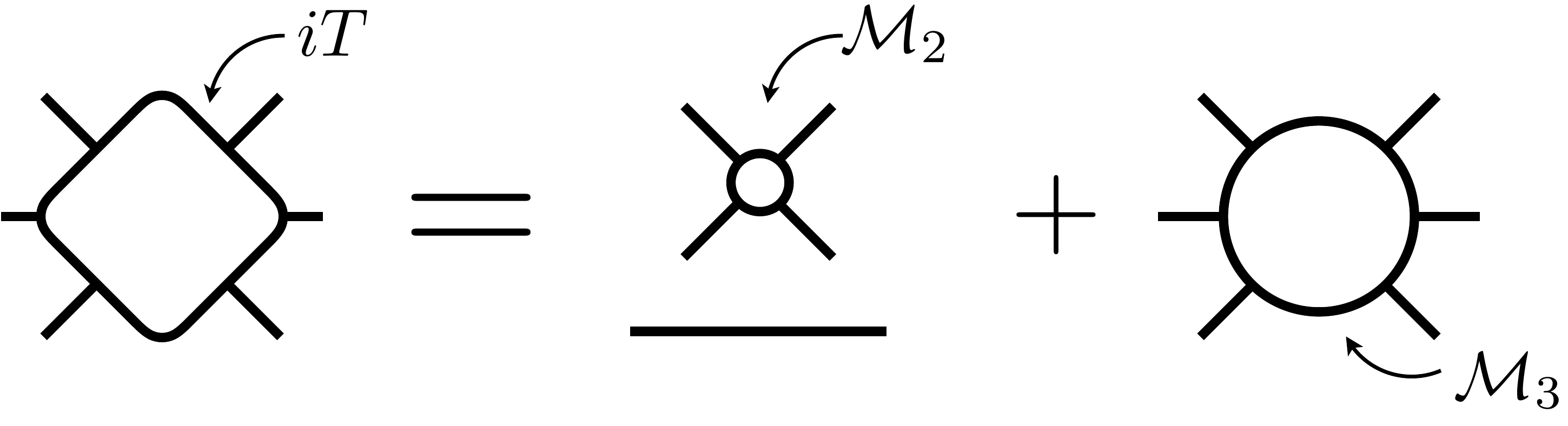}
\caption{ 
The two types of contribution to the three-particle T matrix. $\cM_3$ is the fully-connected amplitude. 
\label{fig:T_mat_def}}
\end{center}
\end{figure}

\subsection{Two-body scattering amplitude}
\label{sec:2body}

The three-body T matrix, illustrated in Fig.~\ref{fig:T_mat_def}, is defined in terms of the S matrix as $iT=S-1$. It has a disconnected contribution, 
depicted as the first term on the right hand side of the 
 figure, in which two particles scatter without interacting with the third, spectator,  particle.   As one would expect, this contribution is fully determined by the two-particle scattering amplitude, denoted
  $\cM_2$. 
To give a useful expression for this, we first define $\vec k$ as the momentum of the spectator particle in some arbitrarily chosen frame. The 4-momentum of this particle is then $(\omega_k, \vec k)$ where $\omega_k = \sqrt{\vec k^2 +m^2}$ and $m$ is the physical mass. We further define the total three-particle energy and momentum in this frame to be $(E, \vec P) \equiv P^\mu$. Thus, if we take one of the incoming scattering particles to carry four-momentum $(\omega_a, \vec a)$, then the second incoming scatterer will have   $(E - \omega_k - \omega_a, \vec P - \vec k - \vec a)$. 

Note that, by enforcing a specific value of total 4-mometum, $(E, \vec P)$, we have given the second scattering particle an energy and momentum that do not necessarily satisfy the on-shell condition. To add this constraint, we need to introduce some new notation. Define
\begin{equation}
E_{2,k}^{\star } \equiv \sqrt{(E - \omega_k)^2 - (\vec P - \vec k)^2} \,,
\end{equation}
as the energy of the two scattering particles in their center-of-mass frame. In other words, the 4-vector $(E - \omega_k , \vec P - \vec k )$ boosts to $(E_{2,k}^\star, \vec 0)$. Denoted by $(\omega_a^\star, \vec a^\star_k)$ is the result of applying this same boost to $(\omega_a, \vec a)$. It then directly follows that $(E - \omega_k - \omega_a, \vec P - \vec k - \vec a)$ is boosted to $(E_{2,k}^\star - \omega_a^\star, -\vec a^\star_k )$. We thus place the third particle on shell by requiring
\begin{equation}
E_{2,k}^\star - \omega_a^\star \overset{!}{=}\omega_a^\star   \ \ \ \Longrightarrow \ \ \ 
a^\star_k \equiv \vert \textbf a^\star_k \vert \overset{!}{=}
 q_{k}^\star \equiv  \sqrt{E_{2,k}^{\star 2}/4 - m^2} \,.
 \label{eq:astarkdef}
\end{equation}

Having enforced this condition we are left with the following (redundant) degrees of freedom for $\cM_2$, viewed as the disconnected contribution to the three-particle T matrix: total 4-momentum [$P^\mu = (E, \vec P)$], spectator momentum ($\vec k$), and incoming and outgoing directional freedom ($\bhat a^\star_k$ and $\bhat a'^\star_k$). This leads us to write the two-particle scattering amplitude as
\begin{align}
\cM_2(\vec k,\bhat a'^\star_k;\vec k,\bhat a^\star_k)
&=
 4\pi Y_{\ell' m'}(\bhat{a}'^\star_k)
\,
\cM_{2; \ell'm';\ell m}(\vec k)
\,Y^*_{\ell m}(\bhat{a}^\star_k)
\,,
\label{eq:M2lm} 
\end{align}
where $Y_{\ell m}$ are 
 the standard spherical harmonics and the sum over repeated angular-momentum indices 
 is implicit. 
We stress that $\bhat{a}^\star_k$ and $\bhat{a}'^\star_k$ are the spherical angles of the relative momenta between the two particles when the system recoils against the same spectator.
%In other words, when $\cM_2$ is embedded in three-to-three scattering, the condition $\vec k = \vec k'$ is imposed. 
 
 In what follows, we will be interested in determining the imaginary contributions to the two- and three-body scattering amplitudes. In doing so, one might have thought that it
would be necessary to keep track of the imaginary parts of the spherical harmonics. Fortunately, it is easy to convince oneself that these contributions exactly vanish. In the case of $\cM_2$ this follows from the fact that,  as a result of Wigner-Eckart theorem, the two-body scattering scattering amplitude is diagonal, and independent of the azimuthal indices $m'$ amd $m$
\begin{align}
\cM_{2; \ell'm';\ell m}(\vec k)
=
\cM^{(\ell)}_{2}(\vec k)
\,\,\delta_{\ell' \ell}\delta_{m' m} \,.
\end{align}
This implies that the product of the two harmonics in Eq.~(\ref{eq:M2lm})  reduces 
to the  real Wigner-$d$ function, or equivalently the Legendre polynomial,
\begin{equation}
\sum_{m = - \ell}^\ell 4\pi \, Y_{\ell m}(\bhat{a}'^\star_k)
\,Y^*_{\ell m}(\bhat{a}^\star_k) =   (2\ell+1) d^{\ell}_{00}  \big (\bhat a'^\star \cdot \bhat a^\star \big ) =  (2\ell+1) P_\ell  \big (\bhat a'^\star \cdot \bhat a^\star \big ) \,,
  \end{equation} 
  for each $\ell$.
%which depends on the cosine   of the relative angle between  $\bhat a'^\star$ and $\bhat a^\star$. 
 An alternative argument is to note that it is legitimate to use real spherical harmonics,
which form a complete set and satisfy the same orthonormality properties as the usual
complex harmonics. Since the harmonics do not appear in the final expressions, we can use
either basis in intermediate steps.
For the real harmonics, the issue with the imaginary part does not arise.
Since all other steps in the derivation have the same form in either basis,
we will get the correct answer if we proceed as if the harmonics are real, even if we use the
complex basis. This argument holds also in the analysis of
the connected three-particle amplitude.

Given that $\cM_{2; \ell'm';\ell m}$ is diagonal, keeping both pairs of indices may seem superfluous. However, as we will see below, it is convenient to think of this as a matrix in angular-momentum space, especially when combining it with other non-diagonal objects. To simplify the notation, 
in what follows we will largely leave the angular-momentum indices implicit. 

\bigskip

We now define the real-valued two-particle K matrix, $\cK_2$, via the standard relation
\begin{equation}
\cM_2(\vec k)^{-1} = \cK_2(\vec k)^{-1} + \rho(\vec k)\,,
\end{equation}
where $\rho$ is imaginary above threshold
\begin{equation}
\label{eq:Imrho}
{\rm Im} \,\rho_{\ell' m';\ell m}(\vec k) = - \delta_{\ell'\ell}\delta_{m'm}\Theta(q_k^{\star2}) 
\bar \rho(\vec k),
\qquad
\bar\rho(\vec k) \equiv \frac{q_{k}^\star}{16 \pi E_{2,k}^\star}\,.
\end{equation}
Here we are using Eqs.~(A6) and (A7) from Ref.~\cite{\HSQCb}, with $\Theta(x)$ the usual Heaviside step function.

From these relations follows
\begin{equation}
{\rm Im}\, \cM_2(\vec k) ={\rm Im} \bigg [ \frac{ \cK_2(\vec k)^{-1} + \rho(\vec k)^*}{ \cK_2(\vec k)^{-1} + \rho(\vec k)^*} \frac{1}{ \cK_2(\vec k)^{-1} + \rho(\vec k)} \bigg ] = \cM_2^*(\vec k) \, \Theta(q_k^{\star2})  \,
\bar\rho(\vec k) \, \cM_2(\vec k) \,,
\label{eq:ImM2}
\end{equation}
where we stress that the overall sign is positive. This result is equivalent to the standard unitarity relation, 
 given in Eq.~(\ref{eq:2bodyUni}) in the Appendix.
 In order to compare to the literature on three-particle amplitudes, 
 and in particular to Ref.~\cite{Jackura:2018xnx}, 
 we note that the exact relation of our $\bar\rho$ to the corresponding quantity in that work is
\begin{equation}
4 \pi \rho^{\text{Ref\,\cite{Jackura:2018xnx}}}_{2k} = 2 \times \bar\rho(\vec k)
\,.
\end{equation}
The factor of $4\pi$ arises because, in Ref.~\cite{Jackura:2018xnx}, the angular integral is left explicit 
[as shown in Eq.~(\ref{eq:unitarity1}) below].
The factor of $2$  arises because of the symmetry factor which reduces the
two-body phase space for identical particles.

We close this subsection by giving an alternative derivation of Eq.~(\ref{eq:ImM2}) that more closely matches the three-particle derivation of the following section. To do so we first expand the relation between $\cM_2$ and $\cK_2$ in powers of the latter
\begin{equation}
\cM_2(\vec k) = \sum_{n=0}^\infty \cK_2(\vec k)  \big [ -  \rho(\vec k) \cK_2(\vec k) ]^n \,.
\label{eq:M2exp}
\end{equation}
To evaluate the imaginary part in this form 
we introduce a general identity for a product of $n$ complex matrices
\begin{equation}
{\rm Im}\,(A_1 A_2 \cdots A_n) = {\rm Im}\,(A_1) A_2 \cdots A_n
+ A_1^*\, {\rm Im}\, (A_2 )A_3 \cdots A_n
+ \cdots + A_1^* \cdots A_{n-1}^* {\rm Im}\, (A_n)
\,.
\label{eq:Impart}
\end{equation}
This follows from simply substituting $2 i\,{\rm Im}\,A_j= A_j-A_j^*$ and noting that terms cancel in pairs. 
The complex conjugation could occur also to the right of the ${\rm Im} (A_j)$ factors, as can be trivially seen by conjugating both sides and using that $\text{Im}(x)$ is real.

Applying this identity to the $n$th term of Eq.~(\ref{eq:M2exp}) then gives
\begin{equation}
{\rm Im} \Big [ \cK_2(\vec k)  \big [ -  \rho(\vec k) \cK_2(\vec k) ]^n \Big ]=  \sum_{m=0}^{n-1}\Big (   \cK_2(\vec k)  \big [ -  \rho(\vec k)^* \cK_2(\vec k) ]^{m}  \Big )    \   \Theta(q_k^{\star2})  \,
\bar\rho(\vec k) \ \Big (       \cK_2(\vec k)  \big [ -  \rho(\vec k) \cK_2(\vec k) ]^{n-m-1}       \Big ) \,,
\end{equation}
where the right-hand side is understood to vanish for $n=0$ since then the sum contains no terms. Summing this result over all $n$ immediately gives Eq.~(\ref{eq:ImM2}). The intuition here is as follows: For a given series of real K matrices and complex valued $\bar\rho$ cuts, the identity (\ref{eq:Impart}) gives a prescription for moving through the chain, summing over all cuts with the conjugated object appearing to the left. Summing over all resulting terms then directly leads to the unitarity relation.

\subsection{Three-body scattering amplitude}
\label{sec:3scatt}

We now turn to the relevant expressions for the fully-connected three-particle scattering amplitude, 
$\cM_3$, which is depicted 
in Fig.~\ref{fig:T_mat_def}.
The three-body scattering amplitude is naturally more complicated than $\cM_2$. In Ref.~\cite{\HSQCb}, 
it was shown in a bottom-up approach based on all-orders perturbation theory, that 
the scattering-amplitude is completely determined  by a real function, denoted $\Kdf$. 
This describes microscopic interactions between the three-particles, i.e.~the part of the scattering amplitude that is not constrained by $s$-channel unitarity.  
For example, in the context of an effective field theory, it is given by a sum of contact interactions and virtual particle exchanges below the three-body threshold~\cite{Bedaque:1998kg,Bedaque:1998km}. 
In the alternative, top-down approaches of Refs.~\cite{
Aaron:1973ca,Amado:1974za,Fleming:1964zz,Aitchison:1966lpz,Mai:2017vot,Jackura:2018xnx},
 one uses S-matrix unitarity to identify the analytic properties and isolate  the analog of $\Kdf$. 
%%%%%%%%%

%%%%%%%%%%%%%%
At this stage, it remains to be shown if the two approaches result in scattering amplitudes with equivalent analytic properties that can be quantitatively matched with a proper choice of the remaining functional freedom. 
 As a first step toward this goal, in this work we 
demonstrate the real-axis unitarity of the  thee-body scattering amplitude,
$\cM_3$, as defined in Ref.~\cite{\HSQCb}. 
In this section we review the result of that work, first by taking the $\Kdf=0$ limit and then by including the all-orders corrections in this short-distance function. With this in hand, in the following section we review the three-body unitarity constraint and show that it is satisfied, order-by-order, by any $\cM_3$ expressed in terms of $\Kdf$.

\subsubsection{Three-body scattering amplitude for $\Kdf=0$}
%%%%%%%%%%%%%%%%%%%%%%%%%%

When $\Kdf=0$, the three-body scattering amplitude is completely determined by pairwise scattering. 
In this case, we have, from Eqs.~(85), (86) and (93) of Ref.~\cite{\HSQCb},
 \begin{equation}
\cM_3(\vec p,\bhat a'^\star_p;\vec k,\bhat a^\star_k) = \cD(\vec p,\bhat a'^\star_p;\vec k,\bhat a^\star_k) \,, \qquad 
\cD(\vec p,\bhat a'^\star_p;\vec k,\bhat a^\star_k) = \cS   \Big \{ \cD^{(u,u)}(\vec p, \vec k)\Big \}\,, \label{eq2} 
\end{equation} 
where $\cD^{(u,u)}$ is the solution to the integral equation
\begin{equation}
i\cD^{(u,u)}(\vec p,\vec k) = i \cM_2(\vec p) iG^\infty(\vec p, \vec k) i \cM_2(\vec k)
+ \int_s i \cM_2(\vec p) iG^\infty(\vec p, \vec s)\, i \cD^{(u,u)}(\vec s,\vec k)
\,.
\label{eq:Duu}
\end{equation}
Note that, since this is a genuine three-particle amplitude, the initial and final momenta,
$\vec k$ and $\vec p$, respectively, differ in general, unlike for $\cM_2$ in Eq.~(\ref{eq:M2lm}).
The objects appearing in Eq.~(\ref{eq:Duu}) are matrices in angular-momentum space, 
with adjacent indices contracted in the usual way. 
The symmetrization operator $\cS$ is defined in Eq.~(37) of Ref.~\cite{\HSQCb} and also explained below, in the paragraph containing Eq.~(\ref{eq:M3symm}).
We use a different shorthand for the integral than in Ref.~\cite{\HSQCb},
namely $\int_s = \int d^3s/[2\omega_s (2\pi)^3]$, with the factor of $\omega_s \equiv \sqrt{\vec s^2 + m^2}$ included. This follows the convention of Ref.~\cite{Jackura:2018xnx}. 

The kinematic function, $G^\infty$, is the pole contribution of the exchange propagator,
defined as
\begin{align}
G^\infty_{\ell' m';\ell m}(\vec p, \vec k)  
&=\left(\frac{k^\star_p}{q_p^\star}\right)^{\ell'}
4\pi Y_{\ell'm'}(\bhat k^\star_p)
\frac{H(\vec p, \vec k)}{u - m^2 + i\epsilon}
Y^*_{\ell m}(\bhat p^\star_k)
\left(\frac{p^\star_k}{q_k^\star}\right)^\ell
\,.
\label{eq:Ginf}
\end{align}
where $u=(P- p- k)^2$.
This is the relativistic form of $G^\infty$, first discussed in Ref.~\cite{\BHSQC}. It differs from 
the nonrelativistic form used in Ref.~\cite{\HSQCb} away from the pole, but all expressions involving
$G^\infty$ remain valid as long as the relativistic form is used throughout.
The function $H(\vec p,\vec k)$ provides a cutoff on $\vec p$ and $\vec k$ and only depends on Lorentz invariant combinations of these momenta with the total momentum $P$. All we need to
know here is that $H$ is real, and that it equals unity when $\vec p$ and $\vec k$ are
chosen so that $u=m^2$.
We re-emphasize  that the magnitudes, $k^\star_p$ and $p^\star_k$, 
entering the angular-momentum barrier factors (as well as $q^\star_k$ and $q^\star_p$),
 are evaluated in the two-particle rest frames, with the subscript giving the spectator momentum.
 They are generalizations of $a^\star_k$ defined in Eq.~(\ref{eq:astarkdef}).
When $\Kdf=0$, the amplitude 
$\cD^{(u,u)}$ is simply the partial-wave projected, unsymmetrized version of ${\cal M}_3$. 

In Eq.~(\ref{eq:Ginf}), $(k^\star_p/q^\star_p)$ and  $(p^\star_k/q^\star_k)$ both equal $1$ at the pole and thus could be omitted from the definition of $G^\infty$ without affecting the properties relevant to unitarity. However, doing so would amount to a redefinition of $\Kdf$ and, 
since these factors cannot be discarded in the finite-volume relation, we prefer to keep them here as well.

We close this subsection by giving more detailed explanations of the possibly unfamiliar notation
used above, so as to make this paper self contained. 
First we relate quantities written in the $\ell m$ basis 
to functions of $\bhat a^\star_k$, all for given spectator-momentum $\vec k$.
This is achieved simply by contracting with spherical harmonics,
as in Eq.~(35) of Ref.~\cite{\HSQCb}. We abuse notation by denoting 
the corresponding quantities using the same symbol, distinguished only by their arguments. 
For example, 
\begin{align}
\cD^{(u,u)}(\vec p,\bhat a'^\star_p;\vec k,\bhat a^\star_k)
&\equiv
4\pi Y_{\ell' m'}(\bhat{a}'^\star_p)
\cD^{(u,u)}_{\ell' m' ; \ell m} (\vec p, \vec k)
\,Y^*_{\ell m}(\bhat{a}^\star_k) \,.
\label{eq:basischange}
\end{align}
In order to lighten the notation, we also sometimes replace the continuous variables $\vec k$ and
$\vec p$ with matrix-like indices, e.g.
\begin{equation}
G^\infty_{p\ell' m';k\ell m} \equiv G^\infty_{\ell' m';\ell m}(\vec p,\vec k)
\,,
\end{equation}
although this does not imply that the spectator momenta are discrete.

\begin{figure}
\begin{center}
\includegraphics[width=.9\textwidth]{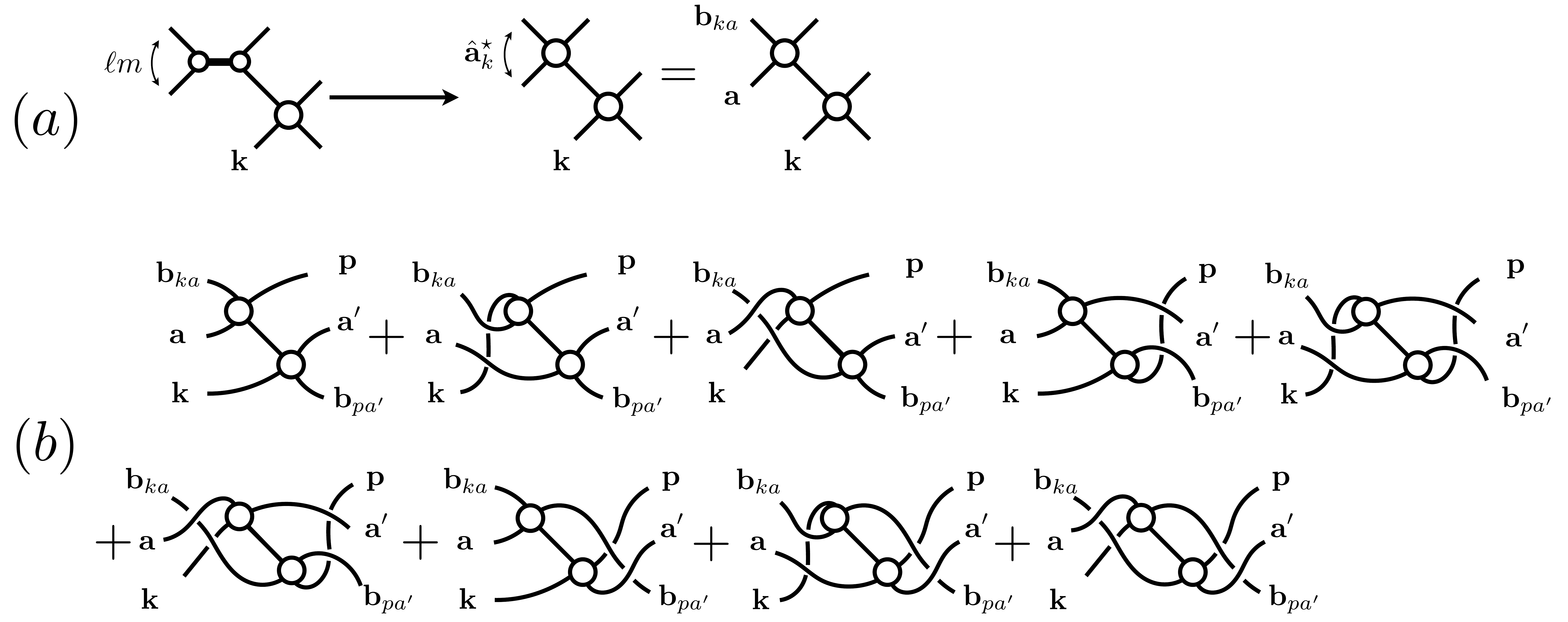}
\caption{
Schematic representation of the definition of symmetrization given
in Eq.~(\ref{eq:M3symm}), using the example of the leading term in $\cD^{(u,u)}$.
(a) Basis transformation from the  $k, \ell,m$ to the on in terms of three momenta; (b) Summing over permutations of assignments of external momenta. Open circles represent $\mathcal M_2$, which is itself symmetric under particle interchange. The third momentum on the left-hand side is given by $\vec b_{ka}= \vec P-\vec k - \vec a$, with $\vec b_{pa'}$ defined similarly. See main text for further explanation.
\label{fig:symm}}
\end{center}
\end{figure}

 Next we recall the definition of symmetrization from Ref.~\cite{\HSQCb}:
\begin{align}
\cM_3(\vec p,\bhat a'^\star_p;\vec k,\bhat a^\star_k)
&=\cS
\left\{
\cM^{(u,u)}_{3; p\ell'm';k\ell m}
\right\}
\equiv
\sum_{x,y=u,s, \tilde{s}}
\cM_3^{(x,y)}(\vec p,\bhat a'^\star_p;\vec k,\bhat a^\star_k)
\,.
\label{eq:M3symm}
\end{align}
Here the superscripts $u$, $s$ and $\tilde{s}$ differ by the choice of spectator momenta.
For example, $\cM_3^{(u,s)}$ is related to $\cM_3^{(u,u)}$ via
\begin{align}
\cM_3^{(u,s)} \big (\vec p,\bhat a'^\star_{p};\vec k,\bhat a^\star_{k} \big )
\equiv
\cM_3^{(u,u)} \big (\vec p,\bhat a'^\star_{p};\vec a,\bhat k^\star_{a} \big ).
\label{eq:us_to_ss}
\end{align}
Note that, in order to symmetrize, we must first change from the $k\ell m$ to the $\vec k, \bhat a^\star_k$
basis. 
The two steps needed to obtain a symmetrized amplitude starting from the
 $k, \ell,m$ basis are summarized in Fig.~\ref{fig:symm},
 for the first term contributing to $\cD(\vec p,\bhat a'^\star_p;\vec k,\bhat a^\star_k)$.
Figure~\ref{fig:symm}(a) represents the  basis transformation, Eq.~(\ref{eq:basischange}),
while Fig.~\ref{fig:symm}(b) shows the nine terms 
that must be summed, each corresponding to the different choices of the 
initial and final spectators.

 We will also need a version of the symmetrization operator,
$\overline\cS$, that acts on objects in the $\vec k, \bhat a^\star$ basis:
\begin{equation}
\overline\cS
\left\{
\cM_3^{(x,y)}(\vec p,\bhat a'^\star_p;\vec k,\bhat a^\star_k)\right\}
\equiv
\sum_{x,y=u,s, \tilde{s}}
\cM_3^{(x,y)}(\vec p,\bhat a'^\star_p;\vec k,\bhat a^\star_k)
\,.\label{eq:Sbar}
\end{equation}

Third, we note that, 
depending on the specific context, either the $ k \ell m$ or the $\vec k, \bhat a^\star$ form
of the amplitudes may be more convenient. For example,
 the first choice is useful in making contact with the finite-volume system, whereas the second choice allows one to better use the exchange symmetry of the underlying amplitudes.
As an example of the latter point,
consider two functions of incoming and outgoing three-particle phase space,
 $\mathcal A$ and $\mathcal B$, assumed to have exchange symmetry. 
 The integrated ``matrix'' product  of the two functions can be expressed in the
following different ways
\begin{align}
\left[\int_s\,
\mathcal{A}(\vec{p},\vec{s})\,\mathcal{B}(\vec{s},\vec{r})
\right]_{\ell_1 m_1;\ell_2 m_2}
&\equiv
\int_s\,
\mathcal{A}_{\ell_1 m_1;\ell m}(\vec{p},\vec{s})\,\mathcal{B}_{\ell m;\ell_2 m_2}(\vec{s},\vec{r})\,,
\\
&=
\left[\int_s
\int_{\bhat{a}^\star_s}
\, \mathcal{A}(\vec{p};\vec{s},\bhat{a}^\star_{s}) \, \mathcal{B}(\vec{s},\bhat{a}^\star_{s};\vec{r})
\right]_{\ell_1 m_1;\ell_2 m_2}\,,
 \\
& =
\int_s
\int_{\bhat{a}^\star_s} \int_{\bhat{a}'^\star_p} \int_{\bhat{a}''^\star_r}\,
4\pi\,
Y^*_{\ell_1m_1}(\bhat{a}'^\star_{p})\,Y_{\ell_2m_2}(\bhat{a}''^\star_{r})\,
\mathcal{A}(\vec{p},\bhat{a}'^\star_{p};\vec{s},\bhat{a}^\star_{s}) \, \mathcal{B}(\vec{s},\bhat{a}^\star_{s};\vec{r},\bhat{a}''^\star_{r})\,,
\label{eq:ang_unit}
\end{align}
where $\int_{\bhat{a}} \equiv \int d\bhat{a}/(4\pi)$.
In the first two lines the exchange symmetry is obscured, whereas in the third it can be directly used via identities such as $\mathcal{A}(\vec{p},\bhat{a}'^\star_{p};\vec{s},\bhat{a}^\star_{s}) = \mathcal{A}(\vec{a}',\bhat{p}^\star_{a'};\vec{a},\bhat{s}^\star_{a})$.
  
\bigskip

Finally, we note that the
expressions given above can be recast in terms of the Lorentz invariants often used in discussions
of three-particle scattering. There are eight independent invariants, usually defined in
the center-of-mass frame, $P = (\sqrt{s},{\bf 0})$.
  A pair of particles is selected in each of the initial and final states and the momentum of the spectator particle corresponds the $\vec k$ and $\vec p$, respectively. 
  The  $\bhat y$  
 axis defines the so-called  production plane and   is given by  $\bhat y = \bhat k\times \bhat p$. The spherical angles, 
 $\bhat a^\star$  and $\bhat a'^\star$ 
  that specify  direction of motion of one of  the two 
   particles in each pair  
    are defined in the respective 
     center-of-mass frame of each pair, with the  $\bhat z$ and $\bhat z'$ axes defined to be 
     opposite to the direction of 
       $\bhat k$ and $\bhat p$ in the two frames, respectively. Note that the $\bhat y$ axis is invariant 
   under boosts from the $P = (\sqrt{s},{\bf 0})$ frame to the rest frames of the two particles subsystems. 
     As for  the     remaining   four variables  
     (besides the two sets of spherical angles $\bhat a^\star$ and $\bhat a^\star$) 
    one can choose the  squares of invariant masses  of the two pairs, $\sigma$ and $\sigma'$ respectively 
    for the initial and final state, the total center of mass energy squared $s=E^2$, 
    and the cosine of the  scattering angle  $z_s= \bhat k\cdot \bhat p$ in the center-of-mass frame.
    As an example of using these invariants, we give the explicit expression for $G^\infty$,
    \begin{align}
G^\infty_{p\ell' m';k\ell m}
=  \sqrt{(2\ell'+1)(2\ell+1)} d^{\ell'}_{m',0}(z_{k}) d^{\ell}_{m,0}(z_{p}) 
 \left(\frac{\lambda(\sigma',u,m^2)}{\lambda(\sigma',m^2,m^2)} \right)^{\ell'/2} 
 \frac{H(\sigma,\sigma') 
  }{u - m^2 + i\epsilon}
  \left(\frac{\lambda(\sigma,u,m^2)}{\lambda(\sigma,m^2,m^2)} \right)^{\ell/2} \,.
\label{eq:Ginf-alt}
\end{align}
where $\lambda(a,b,c) = a^2 + b^2 + c^2 -2 ab- 2ac - 2bc$ is the triangle function,
and the arguments, $z_{k}$ and $z_{p}$ of the Wigner-$d$ functions are cosines of the vectors 
$\bhat k^\star_p $ and $ \bhat p^\star_k$, respectively.
The momentum transfer variable $u$ is given by 
 \begin{equation} 
u = m^2 + \sigma - \frac{(s + m^2 - \sigma') ( s + \sigma - m^2)}{2s} - \frac{\lambda^{1/2}(s,m^2,\sigma)\lambda^{1/2}(s,m^2,\sigma')}{2s} z_s \,.
\end{equation}

\subsubsection{All-orders corrections in $\Kdf$}
%%%%%%%%%%%%%%%%%%%%%%%%%%
 
We now turn to a general expression for the fully-connected three-to-three scattering amplitude.
To do so it is convenient to introduce $\cM_3^{(n)}(\vec p,\bhat a'^\star_p;\vec k,\bhat a^\star_k)$ as the contribution with $n$ powers of the divergence-free K matrix
\begin{align}
\cM_3(\vec p,\bhat a'^\star_p;\vec k,\bhat a^\star_k) &= \cM_3^{(0)}(\vec p,\bhat a'^\star_p;\vec k,\bhat a^\star_k) +
 \cM^{(\mathcal{K})}_{3}(\vec p,\bhat a'^\star_p;\vec k,\bhat a^\star_k)  \,,
\label{eq:DeltaM3}
\\
&= \cM_3^{(0)}(\vec p,\bhat a'^\star_p;\vec k,\bhat a^\star_k) +
\cM^\one_3(\vec p,\bhat a'^\star_p;\vec k,\bhat a^\star_k) +  \cO(\mathcal{K}_{\rm df, 3}^2) \,, 
\label{eq:M3}
\end{align}
where $\cM_3^{(0)}(\vec p,\bhat a'^\star_p;\vec k,\bhat a^\star_k)$ is the contribution considered above
\begin{align}
\cM_3^{(0)}(\vec p,\bhat a'^\star_p;\vec k,\bhat a^\star_k) &=
\cD(\vec p,\bhat a'^\star_p;\vec k,\bhat a^\star_k)\,,
\label{eq:M30}
\end{align}
and $\cM^{(\mathcal{K})}_{3}(\vec p,\bhat a'^\star_p;\vec k,\bhat a^\star_k)$ includes all $\Kdf$ 
dependence with the linear contribution given by 
$\cM_3^{(1)}(\vec p,\bhat a'^\star_p;\vec k,\bhat a^\star_k)$.
As with the $\Kdf$-independent piece, the linear piece is conveniently expressed in terms of its unsymmetrized counterpart
\begin{align}
\cM_3^{(1)}(\vec p,\bhat a'^\star_p;\vec k,\bhat a^\star_k) &=
 \cS\left\{ \cM_{3;p \ell' m'; k \ell m}^{(1,u,u)} \right\}.
\label{eq:M31}
\end{align}

Heuristically this quantity is understood as a single $\Kdf$ insertion dressed with any number of pairwise scatterings on the incoming and outgoing three-particle states. The precise definition is
\begin{align}
 \cM_3^{(1,u,u)}(\vec p, \vec k)  &=
 \int_s\int_r \cL^\uu(\vec p,\vec s) \Kdf(\vec s,\vec r) \cR^\uu(\vec r, \vec k),
\label{eq:M3uu1}
\\
\cL^\uu(\vec p,\vec s) &= \tfrac13 \tilde \delta(\vec p - \vec s) - \cM_2(\vec p) \rho(\vec p)
\tilde \delta(\vec p-\vec s) - \cD^\uu(\vec p, \vec s) \rho(\vec s)\,,
\label{eq:Luu}
\\[5pt]
\cR^\uu(\vec r, \vec k) &= \tfrac13 \tilde \delta(\vec r - \vec k) - \rho(\vec k)\cM_2(\vec k)
\tilde \delta(\vec r-\vec k) -  \rho(\vec r)\cD^\uu(\vec r, \vec k)\,,
\label{eq:Ruu}
\end{align}
where, following Ref.~\cite{Jackura:2018xnx}, we define
\begin{equation}
\tilde \delta(\vec p - \vec k) \equiv (2 \pi)^3 2 \omega_k \delta^3(\vec p - \vec k) \,.
\end{equation}
The delta function in the first terms of $\cL$ and $\cR$ accounts for diagrams with no two-body 
subprocesses. It is accompanied by a factor of $1/3$, which arises because $\Kdf$ is itself a fully symmetric object, meaning   $\mathcal S\big \{ \Kdf \} = 9 \Kdf$ and the factors of $1/3$ cancel this overcounting. 

The all-orders expression for $\cM^{(\mathcal{K},u,u)}_{3}$ can be given by introducing a new quantity, $ \cT$, which 
coincides with $\Kdf$ at leading order and incorporates all higher orders in which all possible pairwise scatterings occur between adjacent short-distance factors. This is encoded in one final integral equation
 \begin{align}
 \cT(\vec p, \vec k) &= \Kdf(\vec p, \vec k) - \int_s\int_r \K_{\df,3}(\vec p, \vec s) \rho(\vec s) \cL^\uu(\vec s, \vec r) \cT(\vec r, \vec k)\,.
 \label{eq:T}
 \end{align}
 The all-orders $\Kdf$-dependent part of $\cM_3$ is then given by
\begin{equation}
 \cM^{(\mathcal{K},u,u)}_{3}(\vec p, \vec k)  =  \sum_{n=1}^\infty \cM^{(n,u,u)}_{3}(\vec p, \vec k)
 = \int_s\int_r \cL^\uu (\vec p, \vec s) \cT (\vec s, \vec r) \cR^\uu (\vec r, \vec k)\,.
 \label{eq:M3uu}
 \end{equation}

\section{Unitarity of the $\Kdf$ to $\cM_3$ relation}
\label{sec:unitarity}

Having reviewed the results of Ref.~\cite{\HSQCb}, we now turn to the main result of this work.
Specifically, in this section we show that any $\cM_3$ satisfying 
 Eqs.~(\ref{eq:DeltaM3}) and (\ref{eq:M3uu}) for real $\Kdf$
 will automatically satisfy the constraints imposed by unitarity. We break the demonstration into three subsections. First we present the constraint (reviewing some details of its derivation in the appendix), then we show that the $\Kdf$-independent piece satisfies unitarity, and finally we demonstrate that this generalizes to the full scattering amplitude, $\cM_3$.

 \subsection{Unitarity constraint for three-body scattering}
%%%%%%%%%%%%%%%%%%%%%%%%%%

To avoid confusion, we label the 
scattering amplitude that emerges from
the top-down unitarity approach by $\mathcal A_3$.
This is logically distinct from the quantity $\cM_3$ that emerges in all-orders perturbation theory through the relation to $\Kdf$. 
We  take $\mathcal A_3$ here as the fully-connected scattering amplitude to make the connection 
to $\cM_3$ as close as possible. As we review in Appendix~\ref{app:Tunitarity}, the disconnected piece 
separately satisfies unitarity.

Unitarity imposes the following constraint on any three-body amplitude for identical particles:
%%%%%%%%%%%%
%%%%%% 
\begin{align}
\begin{split}
{\rm Im}\, \mathcal{A}_3(\bp';\bp) &= 
\frac1{2\times 3!} \int_{p_1''}\int_{p_2''}\int_{p_3''} (2\pi)^4 \delta^4(P-p_1''-p_2''-p_3'')
\mathcal{A}_3^*(\bp';\bp'') \mathcal{A}_3(\bp'';\bp)
\\
&\quad +
\sum_n \Theta(s_n'- 4m^2) {\bar{\rho}(\vec{p}'_n)}
\int \frac{d\bhat a''_{p'_n}}{4\pi} 
\mathcal{A}_{2,nn}^*(\bp';\bp'') \mathcal{A}_3(\bp'';\bp)
\\
&\quad +
\sum_j 
\Theta(s_j-4m^2){\bar{\rho}(\vec{p}_j)} 
\int \frac{d\bhat  a''_{p_j}}{4\pi} 
\mathcal{A}_3^*(\bp';\bp'') \mathcal{A}_{2,jj}(\bp'';\bp)  
\\
&\quad+
\sum_{n,j} \pi \delta(u_{jn}-m^2) \mathcal{A}_{2,n1}^*(\bp';\bp'')
 \mathcal{A}_{2,3j}(\bp'';\bp)
 \,.
 \end{split}
 \label{eq:unitarity1}
\end{align}
We review the derivation of this result in Appendix~\ref{app:Tunitarity}.
The notation here is potentially confusing, so we explain it in detail.
We use a collective notation for the momenta of the three particles, e.g.
 $ \textbf{p}  \equiv \{ \textbf{p}_1,\textbf{p}_2,\textbf{p}_3  \}$. 
The indices $n$ and $j$ each run over the three choices of spectator, 
or equivalently over the possible two-particle subsystems. 
$\mathcal{A}_{2,nj}(\vec p'; \vec p)$ is a two-body scattering amplitude in which
$\bp_j$ and $\bp'_n$ are respectively the spectators in the initial and final states:
 \begin{equation}
 \mathcal A_{2,nj}(\mathbf p'; \mathbf p) = \cM_2(\vec p'_n , \bhat a'^\star_{p'_n} ; 
 \vec p_j ,   \bhat a^\star_{p_j}  ) \,.
 \end{equation}
 This index-heavy notation is needed to accurately specify the last term in Eq.~(\ref{eq:unitarity1}),
as discussed further below. We recall that our notation for $\cM_2$, already introduced in Eq.~(\ref{eq:M2lm}), requires $\vec p'_n =  \vec p_j$. We do not include a delta function in the definition but simply adopt the convention that the amplitude is only written when the vectors are equal and is otherwise ill-defined. 
Finally, we have introduced the Lorentz invariants $s'_n=(P-p'_n)^2$, $s_j=(P-p_j)^2$ and
$u_{jn}=(P-p_j-p'_n)^2$.

We emphasize that Eq.~(\ref{eq:unitarity1}) is closely related to the unitarity condition given 
by Eq.~(8) of Ref.~\cite{Jackura:2018xnx}, 
the only differences being those associated 
with the fact that here we are considering identical particles. 
The first difference is
the need for additional symmetry factors in the first three terms
 [with those in the second and third terms absorbed into the definition of $\bar \rho(\vec k)$].
Next, the sum over $j$ and $n$ in the last term is not constrained. 
This is in contrast to the result of Ref.~\cite{Jackura:2018xnx} where the sum runs over $j \neq n$.
Thus there are nine contributions here rather than six.
The two-index notation for $\mathcal A_2$ is needed here
to encode the fact that two adjacent factors of the two-to-two scattering amplitude cannot arise on the same particle pair. In other words, the spectator of one pairwise scattering must participate in the next. 
Another difference, is that here that $\vec p_j$ simply labels the momentum, 
and not the particle type as it does in Ref.~\cite{Jackura:2018xnx}.
Finally, some kinematic factors have been replaced with $m^2$, due to the simplification of
considering identical particles.

 \subsection{Unitarity of $\cM_3$ when $\Kdf=0$}

We begin by showing that $\cM_3$ satisfies unitarity when $\Kdf=0$. This amounts to evaluating the imaginary part of $\cD^{(u,u)}$ and  showing  that,
after symmetrization, it satisfies Eq.~(\ref{eq:unitarity1}). 
To do so it is convenient to introduce a shorthand in which 
momentum arguments are written as indices,
 while angular momentum indices remain implicit.
Then for example, Eq.~(\ref{eq:Duu}) can be rewritten as
\begin{equation}
\cD^{(u,u)}_{pk} =    -   \cM_{2p} \,  G^\infty_{pk} \, \cM_{2k} 
- \int_s \cM_{2p} \,  G^\infty_{ps} \, \cD^{(u,u)}_{sk}
\,.
\label{eq:Duushort}
\end{equation}

We begin by expanding $\cD^{(u,u)}$ in Eq.~(\ref{eq:Duu}) in powers of $\cM_2$.
Iteratively substituting the expression for $\mathcal D^\uu$ then gives
\begin{equation}
\cD^{(u,u)}_{pk} = \sum_{n=1}^\infty \cD^{(n,u,u)}_{pk}  \,,
\end{equation}
where
\begin{equation}
\cD^{(n,u,u)}_{pk}  \equiv (-1)^n \mathcal M_{2p} \bigg [  \prod_{j=2}^n \int_{s_j} \bigg ]  \prod_{j=1}^n \left(G_{s_j s_{j+1}}^\infty \mathcal M_{2s_{j+1}}\right) \Big \vert_{s_1=p,s_{n+1}=k} \,.
\end{equation}
The notation is cumbersome due to the need to keep track of (and give labels for) the integrated intermediate coordinates. For example, the first three terms are given by
\begin{align}
\cD^{(1,u,u)}_{pk} &  = - \mathcal M_{2p}   G_{pk}^\infty \mathcal M_{2k}  \,, 
\label{eq:D1uu}
\\
\cD^{(2,u,u)}_{pk}  & =   \int_{s_2} \mathcal M_{2p}   G_{ps_2}^\infty \mathcal M_{2s_2} G_{s_2 k}^\infty \mathcal M_{2k} \,, 
\label{eq:D2uu}
\\
\cD^{(3,u,u)}_{pk}  & =  - \int_{s_2}\int_{s_3} \mathcal M_{2p}   G_{ps_2}^\infty \mathcal M_{2s_2} G_{s_2 s_3}^\infty \mathcal M_{2s_3}   G_{s_3 k}^\infty \mathcal M_{2k}   \,.
\label{eq:D3uu}
\end{align}

Our strategy in the following is to build up intuition by showing first how the unitarity condition
is satisfied at quadratic order in $\cM_2$ (i.e.~for $\cD^{(1,u,u)}$), then repeating this analysis
at cubic order (i.e.~including $\cD^{(2,u,u)}$), and finally carrying out the all-orders analysis
by working directly with the integral equation, Eq.~(\ref{eq:Duushort}).

To evaluate the imaginary parts of the various quantities we require a 
compact notation also for the imaginary part of $G^\infty$
\begin{align}
{\rm Im}\, G^\infty_{\ell'm';\ell m}(\vec p, \vec k) & = 
- \Delta_{\ell' m'; \ell m}(\vec p, \vec k)
\,,
\label{eq:ImG} \\
\Delta_{\ell' m'; \ell m}(\vec p, \vec k)
& \equiv
4\pi Y_{\ell' m'} (\bhat k_{p}^\star)\, [  \pi  \delta(b_{pk}^2-m^2) ] \, Y_{\ell m}^*(\bhat p_{k}^\star) \,,
\end{align}
where $b_{pk}^\mu= (E - \omega_p - \omega_k,  \vec P - \vec  p - \vec  k)$.
Note that the $H$ function in Eq.~(\ref{eq:Ginf}) is set to unity by the delta function, since this sets
all three particles on shell. The shorthand version of this result reads $\text{Im} \, G_{pk}^\infty = - \Delta_{pk}$. We also use an abbreviated form of Eq.~(\ref{eq:ImM2}),
$\text{Im}\, \mathcal M_{2k} ^*=  \mathcal M_{2k}^*  \THrho_k     \mathcal M_{2k}  $.

\subsubsection{Unitarity constraint on $\mathcal D^{(1,u,u)}$}

Taking our identity for the imaginary part of a product of matrices, Eq.~(\ref{eq:Impart}), it is now straightforward to evaluate $\text{Im}\,\mathcal D^{(n,u,u)}$.
For example, the leading term gives
\begin{align}
{\rm Im}\, \cD^{(1,u,u)}_{pk} 
=
 \cM^*_{2p} \, \THrho_{p}\, \cD^{(1,u,u)}_{pk}
+
\cD^{(1,u,u)*}_{pk}\,  \THrho_{k} \, \cM_{2 k}
+
\cM^*_{2p} 
\Delta_{pk}
\cM_{2k}
\,.
\end{align}

Before turning to the all orders extension of this, we think it instructive to explain 
how $\mathcal D^{(1,u,u)}$ satisfies Eq.~(\ref{eq:unitarity1}),
 up to terms that contribute at higher orders. 
 To do so we first apply $\mathcal S$ to both sides, to reach
\begin{equation}
\begin{split}
{\rm Im} \, \cD^{(1)}(\vec p, \bhat a''^\star_{p}; \vec k, \bhat a^\star_{k}) &=  
\overline{\mathcal S} \bigg [
\int_{\bhat a'^\star_p}
\cM^*_{2}(\vec p, \bhat a''^\star_{p};\vec p, \bhat a'^\star_p)\,
\Theta(q_p^{\star 2})  \bar\rho(\vec p) \,
 \cD^{(1,u,u)}(\vec p, \bhat a'^\star_p; \vec k,\bhat a^\star_{k})
 \\
 & \hspace{50pt}+
  \,
\int_{\bhat  a'^\star_k}
 \cD^{(1,u,u)*}(\vec p, \bhat a''^\star_{p};\vec k, \bhat a'^\star_k)\,
 \Theta(q_k^{\star2})\bar \rho(\vec k) \,
 \cM_{2}(\vec k, \bhat a'^\star_k; \vec k,\bhat a^\star_{k})
 \\
 & \hspace{50pt}+
\cM^{*}_{2}(\vec p, \bhat a''^\star_{p}; \vec p, \bhat k^\star_{p})
 \,
\pi \delta(b_{pk}^2 - m^2) \,
\cM_2(\vec k, \bhat p^\star_{k}; \vec k, \bhat a^\star_{k}) \bigg ]
\,.
\label{eq:ImD1_symm}
\end{split}
\end{equation}

We see that the last term here exactly corresponds to the last term in Eq.~(\ref{eq:unitarity1}) and the counting is reproduced. The 9 terms in
the symmetrization lead to the 9 terms in the sum over $j$ and $n$.
Thus this term in Eq.~(\ref{eq:unitarity1}) is fully accounted for and there must be no contributions  from higher orders. 
The first term in Eq.~(\ref{eq:ImD1_symm}) leads to a contribution to the second term
in Eq.~(\ref{eq:unitarity1}), in which
$\mathcal A_3 \to \cD^{(1)}$.  The counting here is more tricky: The second term in Eq.~(\ref{eq:unitarity1})
has $3 \times 9 = 27$ contributions of the form $\cM_2\, \cD^{(1,u,u)}$;
3 arise  from the sum over $n$ and 9 from the symmetrization of $\cD^{(1,u,u)}$.
However, of these, only one third, i.e.~9,
 have the $\cM_2 \bar \rho$ attached to $\cD^{(1,u,u)}$ such that the spectators match. 
 Thus only 9 contributions are of the form of the first term in Eq.~(\ref{eq:ImD1_symm}).
 This matches the 9 terms that are obtained when symmetrizing Eq.~(\ref{eq:ImD1_symm}).
 This leaves 18 remaining $\mathcal M_2 \mathcal D^{(1,u,u)}$ type contributions within Eq.~(\ref{eq:unitarity1}) in which the spectators do not match.  
 These arise within $\text{Im}\, \mathcal D^{(2,u,u)}$, and will be identified shortly.
 The same analysis holds for the second term in Eq.~(\ref{eq:ImD1_symm}), which contributes
 to the third term in Eq.~(\ref{eq:unitarity1}).

\subsubsection{Unitarity constraint on $\mathcal D^{(2,u,u)}$}

We next consider $\mathcal D^{(2,u,u)}$. 
From Eq.~(\ref{eq:D2uu}), it is easy to see that imaginary part is
\begin{multline}
{\rm Im}\,
\cD^{(2,u,u)}_{pk} =  
\cM^*_{2p} \,
\THrho_{p}\,
\cD^{(2,u,u)}_{ pk}
+ 
\cD^{(2,u,u)*}_{ pk}\,
\THrho_{k}\,
\cM_{2k} 
\\+
\int_s 
\cM^*_{2p} \,
\Delta_{ps}\,
\cD^{(1,u,u)}_{ sk}
+
\int_s 
\cD^{(1,u,u)*}_{ ps}\,
\Delta_{sk}\,
\cM_{2k} \,
+
\int_s 
\cD^{(1,u,u)*}_{ ps}\,
\THrho_{s}\,
\cD^{(1,u,u)}_{ sk}\,
 \,.
\label{eq:ImD2}                     
\end{multline}
The analysis of the first two terms follows that of ${\rm Im}\,\cD^{(1)}$ and these contribute to
the second and third terms of Eq.~(\ref{eq:unitarity1}), respectively, now with the replacement 
$\mathcal A_3 \to \cD^{(2)}$.
The third and fourth terms in Eq.~(\ref{eq:ImD2}) exactly generate the missing 18 contributions discussed in the preceding paragraph. In other words, one can show that
\begin{align}
\hspace{-15pt}\cS\left\{ \cM^*_{2p} \, \THrho_p\, \cD^{(1,u,u)}_{pk}
 +
 \int_s \cM^*_{2p} \,
 \Delta_{ps}\,
\cD^{(1,u,u)}_{sk}\right\}
& = 
 \sum_j  \Theta \big (q^{\star2}_{p_j} \big) \bar\rho(\vec p_j) 
  \int_{\bhat a^\star_{p_j}}
 \cM_2^*(\bp_j,\bhat a''_{p_j}; \bp_j,\bhat a^\star_{p_j} )
 \cD^{(1)} (\bp_j,\bhat a^\star_{p_j} ; \vec k, \bhat a'^\star_{k})  
 \,,  \label{eq:res1A}
 \\
& = \sum_j \Theta(s_j- 4m^2) {\bar{\rho}(\vec{p}_j)}
\int_{\bhat a''^\star_{p_j}}
\mathcal{A}_{2,jj}^*(\bp;\bp'') \cD^{(1 )}(\bp'';\bp')\Big|_{\bp'_1=\vec k} \,,
 \label{eq:res1B}
 \end{align}
where on the right-hand side of Eq.~(\ref{eq:res1A})
we have used the notation of Ref.~\cite{\HSQCb},  while the second form, Eq.~(\ref{eq:res1B}), 
uses the notation of Ref.~\cite{Jackura:2018xnx} and Eq.~(\ref{eq:unitarity1}).
The additional result needed to show Eq.~(\ref{eq:res1A}) is given by first noting
   \begin{align}
 \sqrt{4\pi}Y^*_{\ell' m'}(\bhat a'^\star_p)
 \int_k \Delta_{\ell' m'; \ell m}(\vec p, \vec k) \mathcal A_{\ell m}^{(u)} (\vec k)
& =
(4\pi)^{3/2} \int_k Y^*_{\ell' m'}(\bhat a'^\star_p)
Y_{\ell' m'} (\bhat k_{p}^\star)\, [  \pi  \delta(b_{pk}^2-m^2) ]
 \, Y^*_{\ell m}(\bhat p_{k}^\star) \mathcal A_{\ell m}^{(u)} (\vec k) \,, 
 \\
& =
4\pi \int_k\delta^2(\bhat a'^\star_p- \bhat k_p^\star)
\, [  \pi  \delta(b_{pk}^2-m^2) ]
 \,  \mathcal A^{(u)} (\vec k,\bhat p_k^\star) \,.
 \end{align}
 Substituting $\mathcal A^{(u)} (\vec k,\bhat p_k^\star)  =   \mathcal A^{(s)} (\vec p,\bhat k_p^\star) $ and evaluating the integral then gives
    \begin{align}
 \sqrt{4\pi}Y^*_{\ell' m'}(\bhat a'^\star_p)
 \int_k \Delta_{\ell' m'; \ell m}(\vec p, \vec k) \mathcal A_{\ell m}^{(u)} (\vec k)
 & =   \Theta(E^{\star 2}_{2,p} - 4 m^2)   \bar \rho(\vec p) 
2 \mathcal A^{(s)}(\vec p,\bhat a'^\star_p) \,,
\\
 & =   \Theta(E^{\star 2}_{2,p} - 4 m^2)   \bar \rho(\vec p) 
\left[ \mathcal A^{(s)}(\vec p,\bhat a'^\star_p) +  \mathcal A^{(\tilde s)}(\vec p,\bhat a'^\star_p) \right]\,.
\label{eq:switchres}
 \end{align}
This holds for any smooth test function that can be decomposed in spherical harmonics
$\mathcal A^{(u)}(\vec k,\bhat a^\star) =Y_{\ell m}^*(\bhat a^\star) \mathcal A_{\ell m}^{(u)}(\vec k)$,
and for which only even values of $\ell$ contribute.
The evenness of $\ell$ follows here from the identical nature of the two particles in the nonspectator
pair. It is needed to obtain the final form, for it allows one to freely replace the superscript $(s)$
with an $(\tilde s)$.
We stress that $\mathcal A^{(s)}(\vec p, \bhat k^*_p)$ is not required to be smooth,
as no harmonic decomposition of this quantity is required in the derivation of the result above.
Thus one can  use the result to show that
\begin{equation}
\int_s \Delta_{ps} \cD^{(1,u,u)}_{sk} =  \THrho_p \left[\cD^{(1,s,u)}_{pk} + \cD^{(1,\tilde s,u)}_{pk} \right]\,,
\label{eq:switchresA}
\end{equation}
despite the fact that $\mathcal D^{(1,s,u)}(\vec p, \bhat a'^\star_p; \vec k, \bhat a^\star_k)$ is singular
in $\bhat a'^\star_p$, and the presence of a similar singularity in $\cD^{(1,\tilde s, u)}$.
Given Eq.~(\ref{eq:switchresA}), one indeed obtains the 18 missing components of the symmetrized 
$\cD^{(1)}$s needed to complete the right-hand sides of Eqs.~(\ref{eq:res1A}) and (\ref{eq:res1B}).

\begin{figure}
\begin{center}
\includegraphics[width=.9\textwidth]{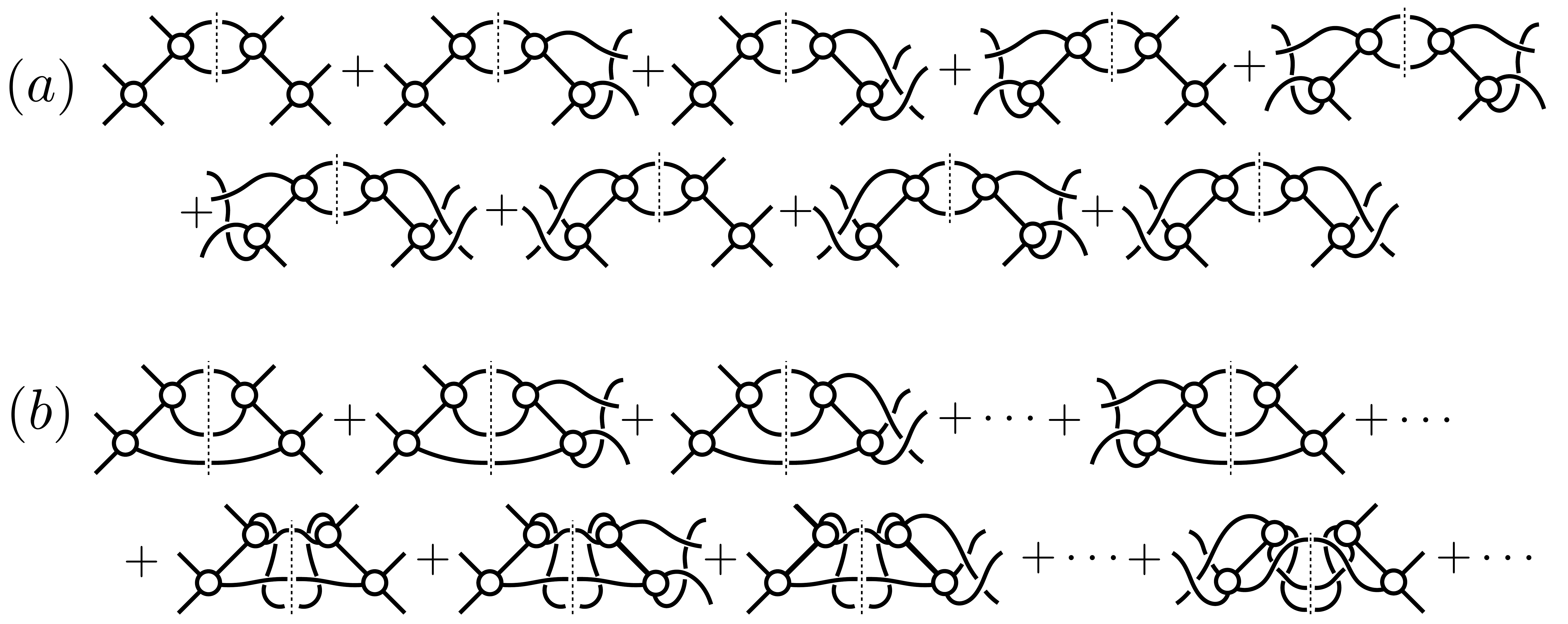}
\caption{
Counting of the $\bar\rho$ contributions in Eq.~(\ref{eq:res2}). (a) The nine contributions from the first term on the left-hand side of Eq.~(\ref{eq:res2}). The notation is as in Fig.~\ref{fig:symm}, except that we do not display the external momenta, which are implicitly held fixed for all diagrams at the same values as in Fig.~\ref{fig:symm}. Dashed lines depict the $\bar\rho$ cut, and the free ends in the middle of the diagram carry the momentum $\vec s$. (b) The $\bar\rho$-type contributions from the right-hand side of Eq.~(\ref{eq:res2}).  Here the dashed line running through all three particles represents the integral on the right-hand side of Eq.~(\ref{eq:res2}). We do not show all 27 contributions; those not displayed are obtained by further permutations of the external lines, 
 following the pattern shown in Fig.~\ref{fig:symm}.
\label{fig:rho_topologies}}
\end{center}
\end{figure}

 Finally, the last term in Eq.~(\ref{eq:ImD2}) gives the first contribution to the first term
 in Eq.~(\ref{eq:unitarity1}). Specifically, we find
\begin{multline}
\cS\left\{ \int_s \cD^{(1,u,u)^*}_{ps} \THrho_s \,\cD^{(1,u,u)}_{sk}
+ \int_s\int_t  \cD^{(1,u,u)^*}_{ps} \Delta_{s t} \cD^{(1,u,u)}_{tk}
\right\}
 = \\
\frac16\frac12 \int_{p_1''}\int_{p_2''}\int_{p_3''} (2\pi)^4 \delta^4(P-p_1''-p_2''-p_3'')
\  \cD^{(1)*} (\bp';\bp'') \  \cD^{(1)} (\bp'';\bp)
\,.
\label{eq:res2}
\end{multline}
Here the situation is similar to that in Eq.~(\ref{eq:res1B}): The first term on the
left-hand side is the symmetrization of the last term in Eq.~(\ref{eq:ImD2}), while
the second term on the left-hand side comes from the next order term, i.e.~${\rm Im} \, \cD^{(3)}$.
Thus one needs only to show that the kinematic and counting factors from the
first term on the left-hand side of Eq.~(\ref{eq:res2}) match those on the right-hand side
coming from the contributions where the $\cM_2$ factors in the two
$\cD^{(1)}$s match.
This correspondence follows directly from
\begin{equation}
\label{eq:twobody}
\int_s \Theta(E_{2,s}^{\star 2} - 4m^2) \bar \rho(\vec s) \int_{\bhat a^\star_s} \mathcal F(\textbf s, \bhat a^\star_s) = 
\frac12 \frac12 \int_{p_1''}\int_{p_2''}\int_{p_3''} (2\pi)^4 \delta^4(P-p_1''-p_2''-p_3'') \mathcal F(\textbf p_1'', \textbf p_2'', \textbf p_3'')  \,,
\end{equation}
where $\mathcal F$ is a test function.
Thus we obtain an overall factor of $1/2$ instead of the required $1/6$.
This is fixed by the relative counting factors. On the left-hand side there
are 9 $\bar \rho$ terms, whereas on the right-hand side there are
$(1/3)\times 9\times 9=27$, as shown in Fig.~\ref{fig:rho_topologies}. The $1/3$ comes from the fact that when joining
two $\cD^{(1)}$s only one third of the terms have the $\bar \rho$ topology---the
others matching with the $\Delta$ term on the left-hand side.
In summary, the counting factors are 9 from the left-hand side and 27 from the right.
These differ by exactly the left over $1/3$ remaining from the $1/6$. 
 
\subsubsection{Unitarity constraint on $\mathcal D^{(u,u)}$}
\label{sec:unitaryD}

We are now ready to argue that $ \mathcal M^{(0)} = \mathcal D$ satisfies the unitarity relation to all orders. 
Starting directly from the integral equation, Eq.~(\ref{eq:Duushort}), and applying the key identity, 
Eq.~(\ref{eq:Impart}), one finds
\begin{multline}
{\rm Im} \cD^{(u,u)}_{pk} = 
- \cM_{2p}^*\, \THrho_p\, \cM_{2p} \, G^\infty_{pk} \, \cM_{2k}
- \cM_{2p}^* \, G^{\infty *}_{pk} \, \cM^*_{2k} \, \THrho_k\, \cM_{2k} 
+ \cM_{2p}^* \, \Delta_{pk} \, \cM_{2k} 
\\
- \int_s  \cM_{2p}^*\, \THrho_p\,  \cM_{2p} \,G^\infty_{ps} \, \cD^{(u,u)}_{sk}
+ \int_s \cM_{2p}^* \,\Delta_{ps} \,\cD^{(u,u)}_{sk}
-\int_s \cM_{2p}^*\, G^{\infty *}_{ps} \,{\rm Im} \cD^{(u,u)}_{sk}
\,.
\label{eq:ImDuu}
\end{multline}
Introducing shorthand for the delta function, $\tilde \delta_{pk} \equiv \tilde \delta (\vec p - \vec k)$, this can be rewritten as
\begin{multline}
\int_s \left(\tilde \delta_{ps} +  \cM_{2p}^* G^{\infty *}_{ps}\right) \, {\rm Im} \cD^{(u,u)}_{sk}
=
\cM^*_{2p} \, \THrho_p\,  \cD^{(u,u)}_{pk}
- \cM_{2p}^*G^{\infty *}_{pk} \, \cM^*_{2k} \, \THrho_k\, \cM_{2k}
 \\ + \cM_{2p}^* \, \Delta_{pk} \, \cM_{2k}
+ \int_s \cM_{2p}^* \, \Delta_{ps} \, \cD^{(u,u)}_{sk}
\,.
\label{eq:ImDuu2}
\end{multline}

\begin{figure}
\begin{center}
\includegraphics[width=.9\textwidth]{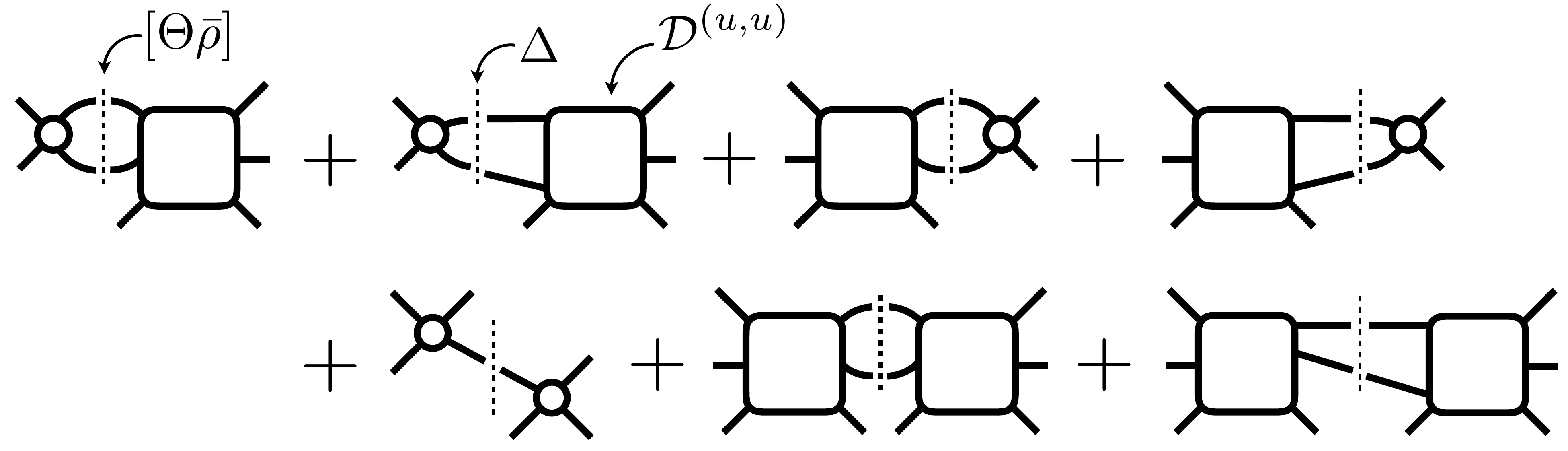}
\caption{
Diagrammatic representation of the right-hand side of Eq.~(\ref{eq:desired}), which gives the imaginary part of $\mathcal D^\uu$. Squares with rounded corners depict $\mathcal D^\uu$, and for simplicity we do not distinguish between $\mathcal D^\uu$ and its conjugate. The two types of cut, defined in the text, are distinguished by the shapes of the adjacent lines.
\label{fig:calD_cuts}}
\end{center}
\end{figure}
At this stage it is helpful to return to the result that we aim to prove, Eq.~(\ref{eq:unitarity1}). Following the intuition developed in the analysis of $\mathcal D^{(1,u,u)}$ and $\mathcal D^{(2,u,u)}$, 
we note that $\mathcal D$ will satisfy the unitarity constraint if $\mathcal D^{(u,u)}$ satisfies
\begin{multline}
{\rm Im}\, \cD^{(u,u)}_{pk} 
=
\cM^*_{2p}\left( \THrho_p \cD^{(u,u)}_{pk}+ \int_s  \Delta_{ps} \cD^{(u,u)}_{sk}\right)
+ \left( \cD^{(u,u)*}_{pk} \THrho_k + \int_s \cD^{(u,u)*}_{ps}\Delta_{sk}\right) \cM_{2k}
\\
 + \cM_{2p}^* \Delta_{pk} \cM_{2k}
+ \left( \int_s \cD^{(u,u)*}_{ps} \THrho_s\, \cD^{(u,u)}_{sk}
+ \int_s\int_t \cD^{(u,u)*}_{ps} \Delta_{st} \cD^{(u,u)}_{tk}\right)
\,.
\label{eq:desired}
\end{multline}
This can be checked by applying $\mathcal S$ to both sides and taking advantage of the internal symmetrizations arising through $\Delta_{pk}$.  We illustrate the right hand side of this equation in Fig.~\ref{fig:calD_cuts}.

Thus our aim is to show that Eq.~(\ref{eq:ImDuu2}) implies Eq.~(\ref{eq:desired}).
What we can easily show, instead, is the opposite implication, namely
that Eq.~(\ref{eq:desired}) implies Eq.~(\ref{eq:ImDuu2}). 
To conclude that the results are in fact equivalent we  
require the additional assumption that the integral operator on the left-hand side 
of Eq.~(\ref{eq:ImDuu2}) is invertible. 
This is plausible since the operator is a deformation of the identity.\footnote{%
In particular, if we discretize the matrix equation, the integral operator becomes the matrix 
$  L^3 \delta^{\text{Kronecker}}_{ps} + \frac{1}{2 \omega_p} \mathcal M_{2p}^* G^\infty_{ps}$. 
This will only have vanishing eigenvalues for specific, fine-tuned choices of discretization, 
encoded here via $L$. Thus we conclude the operator is in general invertible.}
Applying this operator to  Eq.~(\ref{eq:desired}), one finds 
\begin{align}
\int_s\left(\tilde \delta_{ps} + \cM_{2p}^* G^{\infty *}_{ps}\right) {\rm Im}\, \cD^{(u,u)}_{sk}
&= 
\int_s\left(\tilde \delta_{ps} + \cM_{2p}^* G^{\infty *}_{ps}\right) 
\cM^*_{2s}\left( \THrho_s \cD^{(u,u)}_{sk}+ \int_{t}  \Delta_{st} \cD^{(u,u)}_{{t}k}\right)
\nn\\
 &+  
\int_s\left(\tilde \delta_{ps} + \cM_{2p}^* G^{\infty *}_{ps}\right) \left( \cD^{(u,u)*}_{sk} \THrho_k + \int_t \cD^{(u,u)*}_{st}\Delta_{tk}\right) \cM_{2k}
\nn\\
 &+ 
\int_s \left(\tilde \delta_{ps} + \cM_{2p}^* G^{\infty *}_{ps}\right) 
 \cM_{2s}^* \Delta_{sk} \cM_{2k}
\nn\\
 &+  
\int_s\left(\tilde \delta_{ps} + \cM_{2p}^* G^{\infty *}_{ps}\right) 
\left( \int_t \cD^{(u,u)*}_{st} \THrho_t \cD^{(u,u)}_{tk}
+ \int_t\int_l \cD^{(u,u)*}_{st} \Delta_{tl} \cD^{(u,u)}_{lk}\right)
%%%%%%%%%%%%%%%%%%%%%%%%%%%%%%%%%%%%%%%%%%%%%%%%%%%%%%%%%%
\,.
\label{eq:ImD_ugly}
\end{align}

One can simplify this substantially using the integral equation defining
$\cD^{(u,u)}$, Eq.~(\ref{eq:Duu}), which, after complex conjugation and rearrangement,
leads to
\begin{equation}
\int_s\left(\tilde \delta_{ps} + \cM_{2p}^* G^{\infty *}_{ps}\right) \cD^{(u,u)*}_{sk}
= - \cM^*_{2p} G^{\infty *}_{pk} \cM^*_{2k}\,.
\end{equation}
After some straightforward algebra, one finds that the right hand side of Eq.~(\ref{eq:ImD_ugly}) 
equals that of Eq.~(\ref{eq:ImDuu2}). 
 Assuming the invertibility of the integral operator, as discussed above,
it follows that the imaginary part of $\cD^{(u,u)}$ satisfies Eq.~(\ref{eq:desired}),
 and consequently that $\cD$ satisfies unitarity to all orders.

  \subsection{Unitarity of  $\cM_3$ \label{sec:Misunitary}}

Having shown in the previous section that the 
 $\Kdf$-independent part of   $\cM_3$ satisfies Eq.~(\ref{eq:unitarity1}), in this section we show that this holds for the full three-body scattering amplitude. Following the approach of the previous section, we begin with the contribution that is linear in $\Kdf$, denoted $\mathcal M_3^{(1)}$ and then generalize to the full amplitude. 
 
As was the case with $\mathcal D^{(u,u)}$, it is instructive to first determine a constraint equation on $\mathcal M^{(1,u,u)}$ that is equivalent to the general unitarity constraint, Eq.~(\ref{eq:unitarity1})
\begin{multline}
{\rm Im}\, \cM^{(1,u,u)}_{3;pk} 
=
\cM^*_{2p}
\left( \THrho_p \,  \cM^{(1,u,u)}_{3;pk}
+ \int_s  \Delta_{ks}\,\cM^{(1,u,u)}_{3;sk}\,\right)
+ \left( \cM^{(1,u,u)*}_{3;pk}\, \THrho_k 
+ \int_s \cM^{(1,u,u)*}_{3;ps}\, \Delta_{sk}\,\right) \cM_{2k}
\\
+ \left( \int_s \cD^{\uu*}_{ps} \THrho_s \cM^{(1,u,u)}_{3;sk}
+ \int_s\int_t \cD^{\uu*}_{ps} \Delta_{st} \cM^{(1,u,u)}_{3;tk}\right)
+ \left( \int_s \cM^{(1,u,u)*}_{ps} \THrho_s \cD^{\uu}_{3;sk}
+ \int_s\int_t \cM^{(1,u,u)*}_{ps} \, \Delta_{st}\, \cD^{(u,u)}_{3;tk}\right)
\,.
\label{eq:target}
\end{multline} 
If the above is satisfied, $\cM_3$ is consistent with 
 unitarity through first order in $\Kdf$.
The relation is similar in structure to Eq.~(\ref{eq:desired}), 
except that the result here
has more terms because the terms linear in $\Kdf$ can occur both on the left and on the right of the imaginary cut. An analog of the third term in Eq.~(\ref{eq:desired}) is absent here because this term, which leads to the final term in Eq.~(\ref{eq:unitarity1}), is already completely generated by $\mathcal D^{(u,u)}$.
 As with Eq.~(\ref{eq:desired}), the key point is that symmetrizing both sides gives the relevant contribution to the original unitarity constraint.

At this stage we note that our notation is overly complicated for two reasons: 
first, the same combination of $\THrho$ and $\Delta$ appears many times,
and, second, all terms in this, and many of the preceding equations, 
have the form of a matrix product with common indices integrated. 
With this in mind we introduce the shorthand
\begin{equation}
\mathcal I_{ps} = \THrho_p \, \tilde{\delta}_{ps}+  \Delta_{ps} \,,
\label{eq:identSH}
\end{equation}
and to adopt the convention that adjacent factors have a common index 
that is integrated over all values. We emphasize the latter convention by including a dot
wherever there is a common, integrated index.
With this notation, Eq.~(\ref{eq:target}) reduces to
\begin{equation}
{\rm Im}\, \cM^{(1,u,u)}_{3} 
=
(\cM^*_{2}  + \cD^{\uu*}) \cdot \mathcal I \cdot \cM^{(1,u,u)}_{3}
+ \cM^{(1,u,u)*}_{3} \cdot \mathcal I  \cdot (\cM_{2}  + \cD^{\uu})
\,.
\label{eq:target2}
\end{equation}
Note that we are implicitly multiplying $\cM^*_2$ and $\cM_2$ by a delta function in such expressions,
for example
\begin{equation}
\left[\cM^*_2 \cdot \mathcal I\right]_{pr} \equiv \int_t \cM^*_{2p} \tilde{\delta}_{pt} \mathcal I_{tr} = \cM_{2p}^* \mathcal I_{pr} \,,
\end{equation}
with no sum or integral in the final expression.

To show that the result is satisfied, we recall the definition of $\cM_3^{(1,u,u)}$  given in Eqs.~(\ref{eq:M3uu1})-(\ref{eq:Ruu}) above, which in our reduced notation becomes
\begin{gather}
\label{eq:M3uu1B}
 \cM_3^{(1,u,u)}    = \cL^\uu \cdot \Kdf \cdot \cR^\uu \,, 
 \\
\cL^\uu_{ps}  = \tfrac13 \tilde \delta_{ps} - \cM_{2p} \, \rho_p \,
\tilde \delta_{ps} - \cD^\uu_{ps} \, \rho_s \,,
\qquad
\cR^\uu_{rk} = \tfrac13 \tilde \delta_{rk} - \rho_k \, \cM_{2k} \,
\tilde \delta_{rk} -  \rho_r \, \cD^\uu_{rk}\,.
\label{eq:RLdef}
\end{gather}
We begin by taking the imaginary part of $ \cL^\uu_{ps}$
\begin{equation}
{\rm Im}\, \cL^\uu_{ps} = - \cM_{2p}^* \, \THrho_p \,\cM_{2p}\rho_p \tilde{\delta}_{ps}
+ \cM_{2p}^* \, \THrho_p \,\tilde{\delta}_{ps}
- {\rm Im} (\cD^\uu_{ps}) \rho_s
+ \cD^{\uu*}_{ps}\, \THrho_s
\,.
\label{eq:ImLuu1}
\end{equation}
Substituting the result for ${\rm Im} \, \cD^\uu_{ps} $ given by Eq.~(\ref{eq:desired}), we find
\begin{align}
\begin{split}
{\rm Im}\, \cL^\uu_{ps} &=  
\cM_{2p}^* \left\{
\THrho_p\,\tilde{\delta}_{ps}
-\THrho_p \tilde{\delta}_{ps}\,\cM_{2s}\rho_s
-\Delta_{ps} \cM_{2s} \rho_s
- \THrho_p \cD^{(u,u)}_{ps}\rho_s
-\int_t  \Delta_{pt} \cD^{(u,u)}_{ts} \rho_s
 \right\} 
\\
&\hspace{.5cm} 
+ \int_t \cD^{\uu *}_{pt} 
\left\{
\THrho_t \tilde{\delta}_{ts}
 -  \THrho_t \tilde{\delta}_{ts} \cM_{2s}\rho_s
-\Delta_{ts} \cM_{2s}\rho_s
- \THrho_t  \cD^\uu_{ts}\rho_s 
- 
\int_r\Delta_{tr}  \cD^\uu_{rs}\rho_s 
\right\}
\,,
\end{split}
\label{eq:ImLuu2}
\\
&=\int_t\left(\cM^*_{2p}\tilde{\delta}_{pt}+\cD^{\uu *}_{pt} \right)
\left\{\THrho_t \tilde{\delta}_{ts}
-\int_r \mathcal I_{tr} \left(\cM_{2r}\rho_r\tilde{\delta}_{rs}  + \cD^\uu_{rs}\rho_s \right)\right\}\,,
\label{eq:ImLuu2A}
\end{align}
where in the second form we have collected terms by inserting delta-functions as needed
and also by using the definition of $\mathcal I$, Eq.~(\ref{eq:identSH}).
Now we observe that the expression in curly braces in Eq.~(\ref{eq:ImLuu2A}) would
equal $\mathcal I\cdot \cL^{(u,u)}$ were it not for the first term. However, we now make use of 
the replacement identity
\begin{equation}
\label{eq:symmID}
 \THrho_p \, \tilde{\delta}_{ps} \longrightarrow \frac13 \big ( \THrho_p \, \tilde{\delta}_{ps}+  \Delta_{ps} \big ) = \frac{1}{3} \mathcal I_{ps} \,, \qquad {\text{(when\ acting\ on\ a\ symmetric\ object)}}\,.
\end{equation}
This follows from the result Eq.~(\ref{eq:switchres}), since for a symmetric object there is
no difference between versions with $(u)$, $(s)$ and $(\tilde s)$ superscripts.
Note that the result applies irrespective of whether the action on a symmetric object is to the left or
the right.
Here the symmetric object on which $\cL^\uu$ acts is $\Kdf$ (to the right), as can be seen
from Eq.~(\ref{eq:M3uu1B}).

Applying the identity (\ref{eq:symmID}) allows us to write Eq.~(\ref{eq:ImLuu2A}) 
[and its reflection leading to ${\rm Im}\, \cR^\uu$] in a compact form,
\begin{align}
{\rm Im}\, \cL^\uu   & \longrightarrow  
\left(\cM_{2}^* + \cD^{\uu *}\right) \cdot \mathcal I \cdot  \, \cL^{\uu}\,,  
\label{eq:ImLuu3} \\
{\rm Im}\, \cR^\uu  & \longrightarrow  
\cR^{\uu *} \, \cdot \mathcal I \cdot  \left(\cM_{2} +\cD^{\uu}\right)\,.
\label{eq:ImRuu3} 
\end{align}
Since $\Kdf$ is real, we have identified all contributions to ${\rm Im}\,\cM^{(1,u,u)}_{3 } $ [see Eq.~(\ref{eq:M3uu1B})]. We deduce that
\begin{align}
{\rm Im}\,\cM^{(1,u,u)}_{3} &= 
  {\rm Im}\big(\cL^\uu \big) \cdot  \K_{\df,3 } \cdot \cR^\uu 
 +
 \cL^{\uu*} \cdot \K_{\df,3} \cdot {\rm Im}\big(\cR^\uu \big) \,,
\\
&= \left(\cM^*_{2 } + \cD^{\uu*}\right)\cdot \mathcal I \cdot  \cM^{(1,u,u)}_{3 }
+\cM^{(1,u,u)*}_{3 } \cdot \mathcal I \cdot  \left( \cM_{2} + \cD^\uu\right)
\,,
\end{align}
which is indeed the desired result, Eq.~(\ref{eq:target2}). 

\bigskip

Finally, we are in position to demonstrate that the complete connected three-body scattering amplitude, \linebreak
 $\mathcal M_3=\cD + \cM_3^{(\mathcal K)}$, satisfies the unitarity constraint, Eq.~(\ref{eq:unitarity1}).  
 This requires showing that $\cM_3^{(\mathcal K)}$ satisfies the parts of the constraint that
 depend on $\Kdf$. As above, we rewrite these parts as a required constraint
 on the unsymmetrized version of $\cM_3^{(\mathcal K)}$,
\begin{equation}
{\rm Im}\, \cM^{(\mathcal K,u,u)}_{3 } 
=
\left(\cM^*_{2 } + \cD^{\uu*}\right) \cdot \mathcal I \cdot  \cM^{(\mathcal K,u,u)}_{3 }
+ \cM^{(\mathcal K,u,u)*}_{3 } \cdot \mathcal I \cdot  \left(\cM_{2}+\cD^\uu\right)
+ \ \cM^{(\mathcal K,u,u)*}_{3 } \cdot \mathcal I \cdot  \cM^{(\mathcal K,u,u)}_{3}
\,.
\label{eq:targetall}
\end{equation}
Using the same steps as when considering Eqs.~(\ref{eq:target}) and (\ref{eq:target2}),
we find that this symmetrizes to parts of Eq.~(\ref{eq:unitarity1}) that depend on $\Kdf$.
Note that the last term of Eq.~(\ref{eq:unitarity1}) does not need to be produced,
as it is independent of $\Kdf$, and
has already been accounted for by ${\rm Im}\, \cD$.

To demonstrate Eq.~(\ref{eq:targetall}),
we begin by giving the shorthand versions of the results,
 Eqs.~(\ref{eq:T}) and (\ref{eq:M3uu}),
that give the $\Kdf$-dependent part of $\cM_3$, 
 \begin{align}
  \cM^{(\mathcal{K},u,u)}_{3 }  
  = \cL^\uu \cdot \cT \cdot \cR^\uu\,,
\qquad
\cT   = \K_{\df,3 } -  \K_{\df,3 } \cdot \rho \cL^\uu \cdot \cT \,.
 \label{eq:M3uuT}
 \end{align}
The imaginary part of $\cM^{(\mathcal K,u,u)}_{3 }$,
\begin{align}
{\rm Im}\,\cM^{(\mathcal K,u,u)}_{3 } = 
 {\rm Im}\big(\cL^\uu \big) \cdot \cT  \cdot \cR^\uu 
 +
 \cL^{\uu*} \cdot {\rm Im}\big(\cT \big) \cdot \cR^\uu 
 +
   \cL^{\uu*} \cdot \cT  \cdot{\rm Im}\big(\cR^\uu \big) \,,
\end{align}
can be partially evaluated using the replacement rules 
Eqs.~(\ref{eq:ImLuu3}) and (\ref{eq:ImRuu3}),
which can be used since $\cT$ is symmetric.
This leads to
\begin{equation}
{\rm Im}\,\cM^{(\mathcal K,u,u)}_{3 } = 
\left( \cM^*_{2}+  \cD^{\uu*}\right)  \cdot \mathcal I \cdot  \cM^{(\mathcal K,u,u)}_{3 }
+\cM^{(\mathcal K,u,u)*}_3 \cdot \mathcal I \cdot \left( \cM_{2 } +  \cD^\uu  \right)
+ 
 \cL^{\uu*} \cdot {\rm Im}\big(\cT \big) \cdot \cR^\uu 
\,.
\end{equation}
Thus the first two terms in Eq.~(\ref{eq:targetall}) are reproduced,
and all that remains to demonstrate is
\begin{equation}
 \cL^{\uu*} \cdot {\rm Im}\left(\cT \right)  \cdot \cR^\uu 
 =
 \cM^{(\mathcal K,u,u)*}_{3 } \cdot \mathcal I \cdot \cM^{(\mathcal K,u,u)}_{3 }\,.
\label{eq:targetall2}
\end{equation}

To do so requires evaluation of the imaginary part of $\cT$. Using the integral equation 
in Eq.(\ref{eq:M3uuT}),
the result (\ref{eq:ImLuu3}) for the imaginary part of $\cL^\uu$, 
and the reality of $\Kdf$, we find
\begin{equation}
{\rm Im}\,\cT  =  \K_{\df,3 } \cdot \THrho \cL^\uu \cdot \cT 
- \K_{\df,3} \cdot \rho^* (\cM^*_{2} + \cD^{\uu*}) \cdot \mathcal I \cdot \cL^\uu \cdot \cT 
-  \K_{\df,3 } \cdot \rho^* \cL^{\uu*} \cdot {\rm Im}\, \cT 
\,.
 \label{eq:ImT}
 \end{equation}
In the first term on the right-hand side one can apply the symmetrization identity (\ref{eq:symmID}) to write
\begin{equation}
{\rm Im}\,\cT  = 
  \K_{\df,3 } \cdot \cR^{\uu*} \cdot \mathcal I \cdot  \cL^\uu  \cdot \cT 
-    \K_{\df,3 } \cdot \rho^* \cL^{\uu*} \cdot {\rm Im}\, \cT 
\,,
 \label{eq:ImT2}
 \end{equation}
 where we have made use of the definition of $\cR$, given in Eq.~(\ref{eq:RLdef}). 
 The result (\ref{eq:ImT2}) can be rewritten as
 \begin{align}
\mathcal I^{\mathcal K} \cdot   {\rm Im}\,\cT  &=  \K_{\df,3 } \cdot \cR^{\uu*} \cdot \mathcal I \cdot \cL^\uu \cdot \cT  \,,
 \label{eq:ImT3}
 \end{align}
 where we have introduced 
\begin{equation}
\mathcal  I^{\mathcal K}_{pr}  \equiv \tilde{\delta}_{pr}  +  
\int_s \K_{\df,3;ps} \rho_s^* \cL^{\uu*}_{sr}  
   \,,
\end{equation}
which acts as an integral operator.

We now observe that this same operator can be used to rewrite the complex conjugate of
the relation between $\Kdf$ and $\mathcal T$ given in Eq.~(\ref{eq:M3uuT}),  
\begin{equation}
\K_{\df,3 } = \mathcal I^{\mathcal K} \cdot \cT^* \,.
\end{equation}
Thus we find  
\begin{equation}
{\mathcal I}^{\mathcal K} \cdot {\rm Im}\,\cT   =
{\mathcal I}^{\mathcal K} \cdot \cT^*\cdot \cR^{\uu*} \cdot \mathcal I \cdot \cL^\uu \cdot \cT 
\,.
\end{equation}
Assuming that ${\mathcal I}^{\mathcal K}$ is invertible, 
 which is plausible using the same arguments given in Sec.~\ref{sec:unitaryD}
for the integral operator encountered previously,
we can drop the factors of this operator to reach
\begin{equation}
{\rm Im}\,\cT   =
  \cT^* \cdot \cR^{\uu*} \cdot \mathcal I \cdot \cL^\uu \cdot \cT 
\,.
\end{equation}
Finally, inserting this result into the left-hind side of Eq.~(\ref{eq:targetall2}), we immediately
find the right-hand side, concluding the argument.

\section{Conclusion \label{sec:conclusion}}

In this work we have shown that the form of the infinite-volume three-particle scattering amplitude, $\mathcal M_3$, derived in the context of finite-volume formalism, satisfies 
unitarity. Though this result was expected,
the demonstration turns out to be highly nontrivial,
and thus provides an important check of the derivations of Refs.~\cite{\HSQCa,\HSQCb}. 
In particular, the present derivation shows how the factors of $1/3$ in the expressions for
$\cL^\uu$ and $\cR^\uu$ given in Eq.~(\ref{eq:RLdef}) are essential for unitarity to hold. 
Such factors are not present in the alternative  
representations of  Refs.~\cite{Mai:2017vot,Jackura:2018xnx}, and were initially a source of confusion in understanding the consistency of the various approaches.

More generally, we have shown how  
 $\Kdf$ and ${\cal D}$ 
individually  contribute  to the imaginary part of the connected amplitude.   While the latter is a somewhat standard 
 object representing all-orders resummation of the one-particle exchange interactions,  
 $\Kdf$ is a quantity unique to the formalism of Refs.~\cite{\HSQCa,\HSQCb}. It was introduced 
  in Ref.~\cite{\HSQCa} as a  fully symmetric amplitude that encodes the short-distance or microscopic physics. In analogy 
   to the two-particle K matrix it has no unitary branch cuts and is real for real energies. 
   
   In other approaches, e.g.~Refs.~\cite{Mai:2017vot,Jackura:2018xnx},  similar objects appear, 
   but these are not invariant under particle interchange.
   The work presented here can shed light on the connection between these formalisms,
as well as to other approaches that derive three-body amplitudes from unitarity relations~\cite{Amado:1975zz,Aaron:1973ca,Aitchison:1966lpz,Cook:1962zz, BFN:1962, Fleming:1964zz, Holman:1965}.
Indeed, the relation between $\Kdf$ and the $B$-matrix used in Refs.~\cite{Mai:2017vot,Jackura:2018xnx} has been determined in Ref.~\cite{Jackura}.
 We also think that it will be worthwhile investigating the use of the $\Kdf$-parametrization of $\cM_3$
in analyses of experimental data. Given that the finite-volume observables that may be accessed via lattice QCD are more directly related to $\Kdf$~\cite{\HSQCa}, this will serve as a stepping stone towards bridging three-body physics in experiment and lattice QCD.

\section{Acknowledgements}
This work was supported by U.S. Department of Energy contracts DE-SC0011637 (SRS), 
DE-FG02-87ER40365 (APS), and DE-AC05-06OR23177 (RAB, APS) under which Jefferson Science Associates, LLC, manages and operates Jefferson Lab, and  National Science Foundation under Grant No. PHY-1415459 (APS).  RAB also acknowledges support from the U.S. Department of Energy Early Career award, contract DE-SC0019229. SRS also acknowledges partial supported from
the International Research Unit of Advanced Future Studies at Kyoto University.

\appendix

\section{Unitarity relation for the three-body scattering amplitude\label{app:Tunitarity}}
 
In this appendix we review the derivation of Eq.~(\ref{eq:unitarity1}), the constraint that follows from unitarity on the three-particle scattering amplitude.

 Unitarity implies that the $T$ matrix satisfies $T-T^\dag=iT^\dag T$. To derive the resulting constraint
  we evaluate matrix elements of this equation using relativistically-normalized
  three-particle asymptotic states, 
  $|\textbf{p}\rangle =|\textbf{p}_1,\textbf{p}_2,\textbf{p}_3\rangle$.
  This yields 
 \begin{align}
 \langle \textbf{p}'|T|\textbf{p}\rangle
 -\langle \textbf{p}'|T^\dag|\textbf{p}\rangle
 &=
2i{\rm Im}\, \langle \textbf{p}'|T|\textbf{p}\rangle
 =
\frac{i}{3!} \iiint_{p''}
 \langle \textbf{p}'|T^\dag|\textbf{p}''\rangle\,
  \langle \textbf{p}''|T|\textbf{p}\rangle
   =
\frac{i}{3!} \iiint_{p''}
 \langle \textbf{p}'|T|\textbf{p}''\rangle^*\,
  \langle \textbf{p}''|T|\textbf{p}\rangle \,,
  \label{eq:Tunitary0}
 \end{align}
 where we have used
 \begin{equation}
 \langle \textbf{p}' | T^\dagger | \textbf{p} \rangle
 \equiv
 \langle \textbf{p} | T | \textbf{p}' \rangle^*
 =
  \langle \textbf{p}' | T | \textbf{p} \rangle^*
 \,,
 \label{eq:herman}
 \end{equation}
 in which the second equality follows from hermitian analyticity~\cite{EdenHA,OliveHA},
as well as the shorthand notation
 \begin{equation}
 \iiint_p = \int_{p_1} \int_{p_2} \int_{p_3}\,.
\end{equation}
The factor of $3!$ in the denominator is needed for identical particles
to cancel the overcounting arising from integrating over the full three-particle phase space.
The result (\ref{eq:Tunitary0}) can be rewritten as
  \begin{align}
  {\rm Im}\, \langle \textbf{p}'|T|\textbf{p}\rangle
   =
\frac{1}{2\,\times 3!}\int_{p'}
 \langle \textbf{p}'|T|\textbf{p}''\rangle^*\,
  \langle \textbf{p}''|T|\textbf{p}\rangle \,.
  \label{eq:Tunitary}
 \end{align}
 In the following, we also need the total initial and final four-momenta
 \begin{align}
 P^\mu &= (E,\vec P) = (\omega_{p_1}+\omega_{p_2}+\omega_{p_3}, \vec p_1+\vec p_2+\vec p_3)\,,
 \\
P''^\mu &= (E'',\vec P'') = (\omega_{p''_1}+\omega_{p''_2}+\omega_{p''_3}, 
\vec p''_1+\vec p''_2+\vec p''_3)\,.
 \end{align}
 
 Next we decompose the T matrix into disconnected and connected pieces, 
 \begin{equation}
T=T_{\text{d}}+T_{\text{c}}\,,
\label{eq:Tdecomp}
\end{equation} 
with the disconnected piece having the matrix element
  \begin{align}
   \langle \textbf{p}''|T_{\text{d}}|\textbf{p}\rangle
&=
 \sum_{j,k}\tilde{\delta}(\textbf{p}''_j-\textbf{p}_k) 
 (2\pi)^4\delta^4(P''_j-P_k)
 \mathcal{A}_{2,jk}(\textbf{p}'';\textbf{p})\,,
\\
&=  (2\pi)^4\delta^4(P''-P)
 \sum_{j,k}\tilde{\delta}(\textbf{p}''_j-\textbf{p}_k) 
 \mathcal{A}_{2,jk}(\textbf{p}'';\textbf{p})\,,
 \label{eq:Tdisc}
 \end{align}
where the indices $j$ and $k$ run from 1 to 3.
Here $ \mathcal{A}_{2,jk}(\textbf{p}'';\textbf{p})$ is the two-particle scattering amplitude for
the subsystem defined by the fact that the initial and final spectators have momentua $\vec p_k$
and $\vec p''_j$, respectively,
while $P''_j =P-p''_j=(E-\omega_{p''_j},\textbf{P}-\textbf{p}''_j)$ and $P_k=P-p_k$.
The corresponding result for the connected part is
\begin{equation}
   \langle \textbf{p}''|T_{\text{c}}|\textbf{p}\rangle
   =
   (2\pi)^4\delta^4(P'-P)  \mathcal{A}_3(\textbf{p}';\textbf{p})\,,
   \end{equation}
 where $\mathcal A_3$ is the connected three-particle amplitude.

To proceed we insert the decomposition (\ref{eq:Tdecomp}) into the unitarity relation,
Eq.~(\ref{eq:Tunitary}). The left-hand side becomes
\begin{align} 
   {\rm LHS}
&= [{\rm LHS}]_c + [{\rm LHS}]_d\,,
\label{eq:LHS}
\\
[{\rm LHS}]_c &=
    (2\pi)^4\delta^4(P'-P)\, {\rm Im}\, \mathcal{A}_3(\textbf{p}';\textbf{p})\,,
\label{eq:LHSc}
\\
[{\rm LHS}]_d &=
 (2\pi)^4\delta^4(P'-P)    \sum_{j,k}\tilde{\delta}(\textbf{p}'_j-\textbf{p}_k) 
  {\rm Im}\,  \mathcal{A}_{2,jk}(\textbf{p}';\textbf{p}) \,,
\label{eq:LHSd}
 \end{align}
while for the right-hand side we obtain
 \begin{multline} 
   {\rm RHS}
=
\frac{1}{2\,\times 3!}\iiint_{p''}
\bigg\{
 \langle \textbf{p}'|T_{\text{c}}|\textbf{p}''\rangle^*\,
  \langle \textbf{p}''|T_{\text{c}}|\textbf{p}\rangle
  +
 \langle \textbf{p}'|T_{\text{c}}|\textbf{p}''\rangle^*\,
  \langle \textbf{p}''|T_{\text{d}}|\textbf{p}\rangle
 \\
+
 \langle \textbf{p}'|T_{\text{d}}|\textbf{p}''\rangle^*\,
  \langle \textbf{p}''|T_{\text{c}}|\textbf{p}\rangle
  +
 \langle \textbf{p}'|T_{\text{d}}|\textbf{p}''\rangle^*\,
  \langle \textbf{p}''|T_{\text{d}}|\textbf{p}\rangle
  \bigg\}
  \,.
  \label{eq:RHS}
 \end{multline}
 Our aim is to determine the parts of this expression that equal $[{\rm LHS}]_c$,
 for these give the unitarity relation for $\mathcal A_3$.
 
 We label the four terms in Eq.~(\ref{eq:RHS}) as $[{\rm RHS}]_{1-4}$.
 The first three are fully connected, while the last contains disconnected contributions.
 To pull out the latter we
 insert the expression for the disconnected contribution to the T matrix, Eq.~(\ref{eq:Tdisc}),
 into the final term in Eq.~(\ref{eq:RHS}), obtaining
\begin{align}
   [{\rm RHS}]_4
&=\frac{1}{2\,\times 3!}
\sum_{j,k,\ell,n} \iiint_{p''}
\tilde{\delta}(\textbf{p}'_{j}-\textbf{p}''_\ell) 
\tilde{\delta}(\textbf{p}''_{n}-\textbf{p}_{k}) 
 (2\pi)^4\delta^4(P'-P'')
 (2\pi)^4\delta^4(P''-P)
  \mathcal{A}^*_{2,{j\ell}}(\textbf{p}';\textbf{p}'')
  \mathcal{A}_{2,nk}(\textbf{p}'';\textbf{p}).
  \label{eq:RHS4}
\end{align}
Here we are using $j$ and $k$ for the external spectator indices, and $\ell$ and $n$ for the
internal indices. There are two types of contribution to Eq.~(\ref{eq:RHS4}): those in which
$\ell=n$, which are fully disconnected since the same momentum is a spectator for both
scatterings, and the connected contributions in which $\ell\ne n$.
For a given choice of $j$ and $k$, there are 3 contributions of the first kind and 6 of the second.
We denote the fully disconnected contributions by $[{\rm RHS}]_{4d}$, and
the connected by $[{\rm RHS}]_{4c}$.

The three contributions to the fully disconnected part are all equal when considering identical particles,
so we can set $\ell=n=j$ and multiply by an overall factor of three.\footnote{%
The choice of $\ell$ and $n$ does not matter as long as they are equal.
For example, we could equally well choose $\ell=n=k$ or $\ell=n=1$.}
This allows us to write 
\begin{align}
   [{\rm RHS}]_{4d}
&=\frac{ (2\pi)^4\delta^4(P'-P) }{2\,\times 2}
\sum_{j,k} \iiint_{p''} \tilde{\delta}(\textbf{p}'_{j}-\textbf{p}_k)
\tilde{\delta}(\textbf{p}''_{j}-\textbf{p}_{k}) 
 (2\pi)^4\delta^4(P''-P)
  \mathcal{A}^*_{2,{jj}}(\textbf{p}';\textbf{p}'')
  \mathcal{A}_{2,jk}(\textbf{p}'';\textbf{p}) \,.
  \label{eq:RHS4d}
\end{align}
Next we use the result Eq.~(\ref{eq:twobody}), which, after carrying
 out the remaining spectator-momentum integral, gives
\begin{align}
 [{\rm RHS}]_{4{\text{d}}}
&=
{ (2\pi)^4\delta^4(P'-P)}
\sum_{j,k}
 \Theta(E_{2,p_k}^{*\, 2} - 4 m^2) \bar\rho(\vec{p}_k)
 \tilde{\delta}(\textbf{p}'_j-\textbf{p}_{k}) 
 \int \frac{d \bhat a^\star_{p''_j}}{4\pi}
\mathcal{A}^*_{2,jj}(\textbf{p}';\textbf{p}'')
  \mathcal{A}_{2,jk}(\textbf{p}'';\textbf{p})\,.
  \label{eq:RHS4dA}
  \end{align}
We recall that $\bhat a^\star_{p''_j}$ is the direction of one of the intermediate particles in the 
center of mass of the two-particle subsystem for which $\vec p''_j$ is the spectator momentum.

Equating the fully disconnected contributions to the left- and right-hand sides of the unitarity relation,
which are given respectively by Eqs.~(\ref{eq:LHSd}) and (\ref{eq:RHS4dA}),
we find 
\begin{align}
    {\rm Im}\, 
 \mathcal{A}_{2,jk}(\textbf{p}';\textbf{p})
&=
\Theta(E_{2,p_k}^{*\, 2} - 4 m^2) \bar\rho(\vec{p}_k) 
 \int \frac{d \bhat a^\star_{p''_j}}{4\pi}
\mathcal{A}^*_{2,jj}(\textbf{p}';\textbf{p}'')
  \mathcal{A}_{2,jk}(\textbf{p}'';\textbf{p}) \,.
  \label{eq:2bodyUni}
  \end{align}
This is  the standard unitarity constraint on the two-body scattering amplitude,
as given, for example, in Eqs.~(6) and (7) of Ref.~\cite{Jackura:2018xnx}, taking into account that
here we have an additional factor of $1/2$ on the right-hand side due to our use of identical particles.

We now evaluate the connected piece of Eq.~(\ref{eq:RHS4}),
\begin{align}
\hspace{-10pt}   [{\rm RHS}]_{4\text{c}}
&=\frac{ (2\pi)^4\delta^4(P'-P)}{2\,\times 3!}
\sum_{j,k,\ell\neq n}
\iiint_{p''} 
\tilde{\delta}(\textbf{p}'_{j}-\textbf{p}''_\ell) 
\tilde{\delta}(\textbf{p}''_{n}-\textbf{p}_{k}) 
 (2\pi)^4\delta^4(P''-P)
  \mathcal{A}^*_{2,{j\ell}}(\textbf{p}';\textbf{p}'')
  \mathcal{A}_{2,nk}(\textbf{p}'';\textbf{p})
\,, \\
&=\frac{ (2\pi)^4\delta^4(P'-P)}{2}
\sum_{j,k}
\iiint_{p''} 
\tilde{\delta}(\textbf{p}'_{j}-\textbf{p}''_{\bar k}) 
\tilde{\delta}(\textbf{p}''_{\bar \jmath}-\textbf{p}_{k}) 
 (2\pi)^4\delta^4(P''-P)
  \mathcal{A}^*_{2,{j 1}}(\textbf{p}';\textbf{p}'')
  \mathcal{A}_{2,3 k}(\textbf{p}'';\textbf{p})
    \,, \\
  &={ (2\pi)^4\delta^4(P'-P)}
\sum_{j,k}
\pi\,\delta(b_{jk}^2-m^2)\,
  \mathcal{A}^*_{2,{j 1}}(\textbf{p}';\textbf{p}'')
  \mathcal{A}_{2, 3 k}(\textbf{p}'';\textbf{p}) \,.
\end{align}
To obtain the second line we have used the fact that all six terms in the sum over $\ell\ne n$ 
are equal, since they differ only by the choice of dummy indices. We have thus made a canonical
choice ($\ell=1$ and $n=3$) and multiplied by six. 
To obtain the final line we have simply carried out the integrals, and used the
definition $b_{jk}\equiv P-p'_j-p_k$.

The remaining terms in Eq.~(\ref{eq:RHS}) are more straightforward to evaluate.
 For example, the second term gives
 \begin{align} 
   {\rm [ RHS]_2}
&=
  \frac{1}{2\,\times 3!}
   \sum_{jk}
   \iiint_{p''} 
    \langle \textbf{p}'|T_{\text{c}}|\textbf{p}''\rangle^*\,
\tilde{\delta}(\textbf{p}''_j-\textbf{p}_k) 
 (2\pi)^4\delta^4(P''-P)
 \mathcal{A}_{2,jk}(\textbf{p}'';\textbf{p}) \,,
  \\
  &=
    \frac{ (2\pi)^4\delta^4(P'-P)}{2\,\times 3!}
   \sum_{jk}
   \iiint_{p''} 
\mathcal{A}^*_3(\textbf{p}';\textbf{p}'')\,
\tilde{\delta}(\textbf{p}''_j-\textbf{p}_k) 
 (2\pi)^4\delta^4(P''-P)
 \mathcal{A}_{2,jk}(\textbf{p}'';\textbf{p}) \,,
  \\
  &=
  { (2\pi)^4\delta^4(P'-P)}
   \sum_{j}
\Theta(E_{2,p_j}^{*\, 2} - 4 m^2)   \bar \rho(\vec{p}_j) 
 \int \frac{d \bhat a''^\star_{p_j}}{4\pi}
\mathcal{A}^*_3(\textbf{p}';\textbf{p}'')\,
 \mathcal{A}_{2,jj}(\textbf{p}'';\textbf{p}) \,.
  \label{eq:RHS2}
 \end{align}
  In going from the second to third line, we have again used Eq.~(\ref{eq:twobody}),
   and the fact that we are summing over three values of $k$, all of which give the same contribution.
   
    The third term in Eq.~(\ref{eq:RHS}) can be written similarly 
    (interchanging $\vec p \leftrightarrow \vec p'$ and complex conjugating),
    while the first term becomes
    \begin{equation}
    {\rm [RHS]_1} = 
 \frac{ (2\pi)^4\delta^4(P'-P)}{2 \times 3!} \iiint_{p''}   { (2\pi)^4\delta^4(P''-P)}
  \mathcal A^*_3(\vec p';\vec p'') \mathcal A_3(\vec p'';\vec p)\,.
    \label{eq:RHS1}
    \end{equation}
Equating $[{\rm LHS}]_c$ to ${\rm [RHS]_1} + {\rm [RHS]_2} + {\rm [RHS]_3} +  {\rm [RHS]_{4c}}$
leads to the claimed unitarity relation, Eq.~(\ref{eq:unitarity1}).

\bibliography{ref} %%% ref.bib file

\end{document}